\documentclass[structabstract]{aa} 
\newcommand{\kms}{$\mathrm{km\, s^{-1}\, }$}
\newcommand{\msun}{M_{\odot}}

\usepackage{natbib}
\usepackage{graphicx}
\usepackage{longtable}
\usepackage{amsmath} 

\usepackage{txfonts}
\begin{document}
   \title{$N$-body simulations of globular clusters in tidal fields: \\ Effects of intermediate-mass black holes.}
   
   \titlerunning{$N$-body simulations of globular clusters in tidal fields}

   \author{N. L\"utzgendorf
          \inst{1}
          \and
          H. Baumgardt \inst{2}
          \and
          J. M. D. Kruijssen \inst{3}              
          }

   \institute{European Southern Observatory (ESO),
              Karl-Schwarzschild-Strasse 2, 85748 Garching, Germany\\
              \email{nluetzge@eso.org}
         \and
			 School of Mathematics and Physics, University of Queensland, 
			 Brisbane, QLD 4072, Australia
		\and
			 Max-Planck Institut f\"ur Astrophysik,
             Karl-Schwarzschild-Stra\ss e 1, D-85748
             Garching, Germany 
}

   \date{Received May 21, 2013; accepted August 05, 2013}

 
  \abstract
   {Intermediate-mass black holes (IMBHs) may provide the missing link to understanding the growth of supermassive black holes in the early Universe. Some formation scenarios predict that IMBHs could have formed by runaway collisions in globular clusters (GCs). However, it is challenging to set observational constraints on the mass of a black hole in a largely gas-free (and hence accretion-free) stellar system such as a GC. Understanding the influence of an IMBH in the center of a GC on its environment might provide indirect detection methods.}
   {Our goal is to test the effects of different initial compositions of GCs on their evolution in a tidal field. We pin down the crucial observables that indicate the presence of an IMBH at the center of the cluster. In addition to central IMBHs, we also consider the effects of different stellar-mass black hole retention and stellar binary fractions.}
   {We performed a set of 22 $N$-body simulations and varied particle numbers, IMBH masses, stellar-mass black-hole retention fractions, and stellar binary fractions. These models are all run in an external tidal field to study the effect of black holes on the cluster mass loss, mass function, and life times. Finally, we compared our results with observational data. }
   {We found that a central massive black hole increases the escape rate of high-mass stars from a star cluster, implying that the relative depletion of the mass function at the low-mass end proceeds less rapidly. Furthermore, we found a similar behavior for a cluster hosting a high number of stellar-mass black holes instead of one massive central IMBH. The presence of an IMBH also weakly affects the fraction of the cluster mass that is constituted by stellar remnants, as does the presence of primordial binaries. We compared our simulations with observational data from the literature and found good agreement between our models and observed mass functions and structural parameters of GCs. We exploited this agreement to identify GCs that could potentially host IMBHs.}
{}


   \maketitle
%

\section{Introduction}

The dynamical evolution of globular clusters (GCs) in a tidal field causes them to gradually dissolve \citep[e.g.][]{ambartsumian38,spitzer58,baumgardt_2001}. After GCs have migrated out of their gas-rich birth environment \citep{shapiro_2010,elmegreen_2010,kruijssen_2012}, this dissolution is driven by two-body relaxation, that is, the repeated effect of soft encounters between stars, which leads to energy equipartition and populates the high-velocity tail of the Maxwellian velocity distribution \citep[e.g.][]{portegieszwart10}. Because escape is driven by a tendency toward energy equipartition between stars of different masses and a mass-dependent encounter rate, the escape probability is not the same for all stars and generally decreases with mass \citep{henon69,vesperini97b,takahashi00,baumgardt_2003,kruijssen_2009}. As a result of mass segregation, high-mass stars sink to the center of a star cluster, while low-mass stars are pushed to the outskirts, where they are easily removed by the Galactic tidal field. This changes the slope of the mass function (MF) and the mass-to-light ratio (M/L) of the GC \citep[e.g.][]{richer91,demarchi_2007,mandushev91,kruijssen09,baumgardt_2003}.

How exactly the content of evaporating clusters evolves depends on their initial composition and global properties. \citet{kruijssen_2009} showed that the most massive objects in a cluster play a key role in determining which bodies escape. It is therefore important at which point the bulk of the dynamical mass loss occurs, that is, early on (when massive stars are still present) or after $\sim1$~Gyr (when all stars with masses $m>2~\msun$ will have died). In the latter case, the cluster evolution will be dominated by stellar remnants, because at such old ages they generally have masses that exceed the masses of the most massive luminous stars. During their formation process, remnants receive velocity kicks \citep[e.g.][]{lyne_1994,pfahl02,moody09,fregeau09}, and hence the fraction of remnants that is retained depends on the depth of the potential well $M/r_{\rm h}$ (see \citealt{kruijssen_2009} for a quantitative analysis), as well as on the magnitude of the kick velocity.

If the central density of a young GC is high enough, runaway collisions between stars may be able to produce an intermediate-mass black hole \citep[IMBH, see][]{portegies_2004,gurkan_2004}. However, recent studies have shown \citep[e.g.][]{glebbeek_2009} that stellar mass-loss complicates the formation of intermediate-mass black holes through runaway collisions. In the context of the above discussion, it is clear that the presence of an IMBH could potentially have an observable influence on the evolution of the GC composition. Thus far, the search for IMBHs has focused on direct detection through kinematics \citep[e.g.][]{baumgardt_2003a,bosch_2006,noyola_2008,nora11}, surface brightness profiles \citep[e.g.][]{noyola_2006}, or the search for Bondi-Hoyle accretion in the centers of GCs \citep[e.g.][]{strader12}. An analysis of how the presence of IMBHs changes the stellar composition of GCs would provide an independent tracer, or at the very least facilitate obtaining better criteria for the selection of target GCs in observational work.

Previous numerical work on IMBHs in GCs has had a natural emphasis on the internal evolution of GCs. \cite{baumgardt_2005} and \cite{noyola_2011} found that the surface brightness profiles of clusters hosting an IMBH exhibit weak central cusps, in contrast to core-collapsed clusters with very steep profiles and pre-core collapsed systems with no cusp at all. However, \citet{trenti_2010} and \citet{vesperini_2010} showed that the shallow cusp may also form as a transition state of GCs undergoing core collapse and therefore cannot be a sufficient criterion for a GC hosting an IMBH. Furthermore, it has been shown that IMBHs prevent a cluster from undergoing  core collapse and reduce the degree of mass segregation compared with non-IMBH clusters \citep{baumgardt_2004b,gill_2008}. \cite{trenti_2007} predicted that clusters with a high ratio of core radius to half-mass radius are good candidates for hosting an IMBH. This was challenged by \cite{hurley_2007}, who showed that the ratios observed for Galactic GCs can be explained without the need for an IMBH when treating model data as if they were observational data. Most of these studies focused on internal observables such as the surface brightness profile and hence did not consider tidal effects. Because of the importance of a tidal boundary for the eventual composition of a GC, our simulations do include a tidal field. A first effort in this direction was made by \cite{trenti_2010}, who included a single $N$-body run with an IMBH in their model grid of tidally dissolving clusters.

 \begin{figure*}
  \centering
   \includegraphics[width=\textwidth]{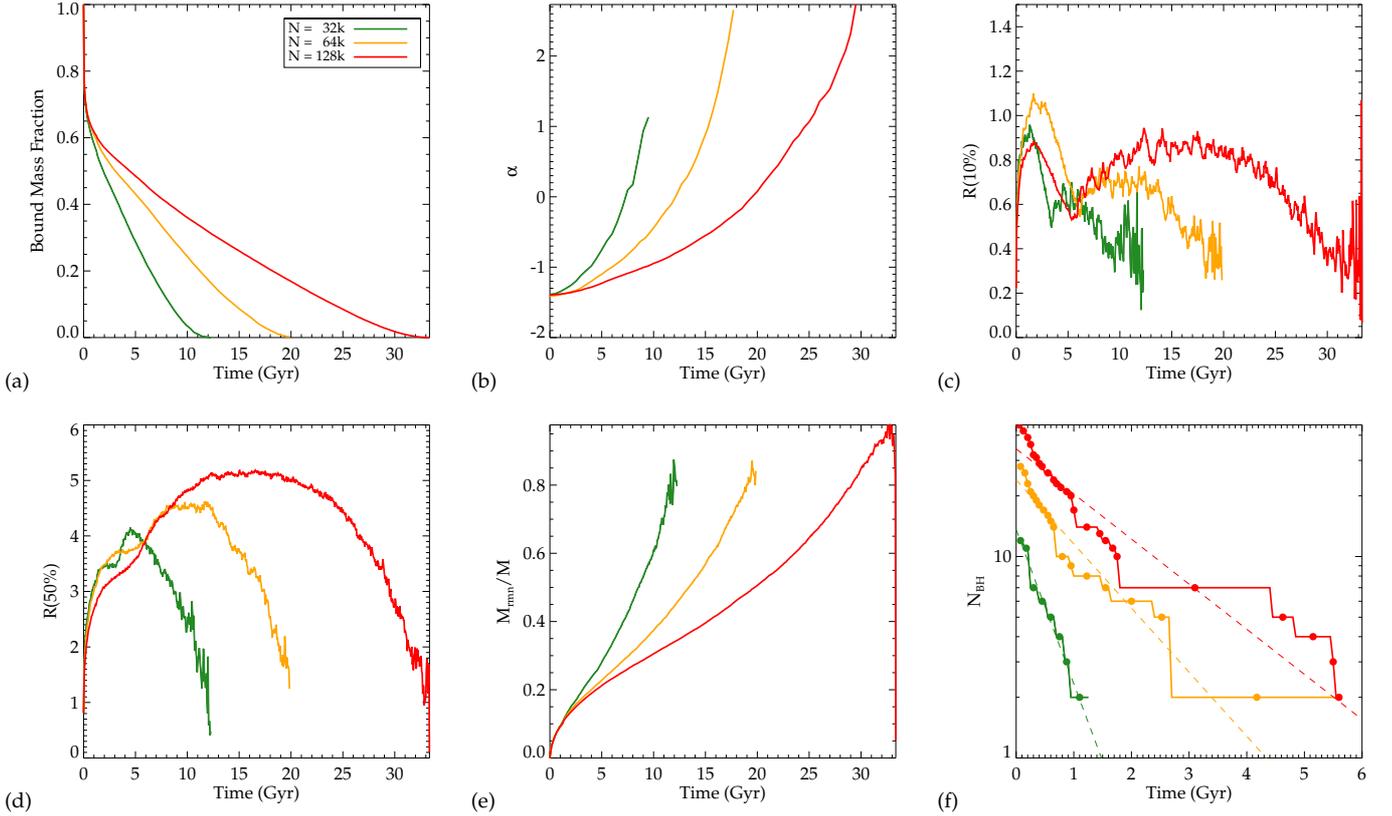}
      \caption{Evolution of the reference models. Shown are a) the bound mass fraction, b) the slope of the mass function $\alpha$ (for $0.3 \msun <  m < 0.8 \msun$), c) the Lagrangian radii $r_{10\%}$ and d) $r_{50\%}$, e) the remnant mass fraction, and f) the number of stellar-mass black holes as a function of time. We stress that the absolute time on the x-axis is arbitrary because it depends on the total mass and the galactocentric distance and is therefore only shown for relative comparison. Dashed lines in panel f) mark the logarithmic fits to the evolution of the black hole numbers.}
         \label{fig:ref}
   \end{figure*}

In this paper, we address the dynamical evolution of GCs with direct $N$-body simulations, serving two particular aims. Firstly, we study the evolution of the composition of dissolving clusters itself, covering a substantial part of the parameter space. In doing so, we vary quantities that are important but have not received much prior attention, such as the remnant retention fraction and the initial binary fraction. Secondly, we intend to identify global tracers of the presence of IMBHs or massive remnants.  We discuss the evolution of the characteristics of GCs and compare them with observational results. Wherever possible, we identify potential targets for future observational searches for IMBHs.

The structure of the paper is as follows: In Section \ref{sec:mod}, we introduce the $N$-body code and the initial conditions of the simulations. Section \ref{sec:obs} is dedicated to the influence of different IMBH masses, black-hole retention fractions, and binary fractions on the properties of tidally dissolving clusters. In Section \ref{sec:comp}, we compare our numerical results with the observed distribution of GCs in the mass function slope versus concentration plane. The paper concludes in Section \ref{sec:con}, where we summarize our work and present our conclusions. 


\section{Models}\label{sec:mod}

In this section we describe the survey of $N$-body simulations that are treated in this paper.

\subsection{$N$-body simulations}

We ran $N$-body simulations based on the GPU (Graphic Processing Unit)-enabled version of the collisional $N$-body code NBODY6 \citep{aarseth_1999,nitadori_2012} on GPU graphic cards at the Headquarters of the European Southern Observatory (ESO) in Garching and the University of Queensland in Brisbane. This code uses a Hermite integration scheme with variable time steps. Furthermore, it treats close encounters between stars by applying KS \citep{KS} and chain regularizations and accounts for stellar evolution \citep{hurley_2000}. The regularization procedures are crucial for following orbits of tightly bound binaries over a cluster lifetime accurately and treat strong binary-single and binary-binary interactions properly. The simulations were carried out with particle numbers of $N=32~768$ (32k), $65~536$ (64k), and $131~072$ (128k) stars.

\begin{table}
\caption{Initial parameters of the $N$-body simulations. The similar masses of some of the models are due to an error in the randomizing routine but do not affect the individuality of each model.}             
\label{tab:start}      
\centering
\begin{tabular}{llllll}
\hline \hline
\noalign{\smallskip}
\multicolumn{1}{c}{$N$} & \multicolumn{1}{c}{$M_{\rm{0}}$} & \multicolumn{1}{c}{$M_{\bullet}/M_{\rm{0}}$} & \multicolumn{1}{c}{$f_{\rm{ret}}$}& \multicolumn{1}{c}{$f_{\rm{bin}}$} & \multicolumn{1}{c}{$T_{\rm{diss}}$} \\
    & \multicolumn{1}{c}{$(M_{\odot})$} & & & & \multicolumn{1}{c}{(Gyr)} \\
\noalign{\smallskip}
\hline
\noalign{\smallskip}
\multicolumn{6}{c}{\tiny{- REFERENCE MODELS -}} \\
\noalign{\smallskip}
$32768$........... & $20887$ & $0.00$ & $0.30$ & $0.00$ & $9.55$ \\
$65536$........... & $41678$ & $0.00$ & $0.30$ & $0.00$ & $16.55$  \\
$131072$......... & $84050$ & $0.00$ & $0.30$ & $0.00$ & $27.35$  \\
\noalign{\smallskip}
\multicolumn{6}{c}{\tiny{- FAMILY 1: IMBH MASSES -}} \\
\noalign{\smallskip}
$32768$........... & $20686$ & $0.01$ & $0.30$ & $0.00$ & $9.25$  \\
$65536$........... & $41678$ & $0.01$ & $0.30$ & $0.00$ & $16.20$  \\
$131072$......... & $84050$ & $0.01$ & $0.30$ & $0.00$ & $28.25$  \\
\noalign{\smallskip}
$32768$........... & $20686$ & $0.03$ & $0.30$ & $0.00$ & $7.90$  \\
$65536$........... & $41678$ & $0.03$ & $0.30$ & $0.00$ & $13.80$  \\
$131072$......... & $84000$ & $0.03$ & $0.30$ & $0.00$ & $24.50$  \\
\noalign{\smallskip}
\multicolumn{6}{c}{\tiny{- FAMILY 2: BH RETENTION FRACTION -}} \\
\noalign{\smallskip}
$32768$........... & $20686$ & $0.00$ & $0.00$ & $0.00$ & $9.10$  \\
$65536$........... & $41678$ & $0.00$ & $0.00$ & $0.00$ & $15.45$  \\
$131072$......... & $83037$ & $0.00$ & $0.00$ & $0.00$ & $26.60$  \\
\noalign{\smallskip}
$32768$........... & $20686$ & $0.00$ & $0.50$ & $0.00$ & $10.10$  \\
$65536$........... & $41678$ & $0.00$ & $0.50$ & $0.00$ & $16.55$  \\
$131072$......... & $83754$ & $0.00$ & $0.50$ & $0.00$ & $28.35$  \\
\noalign{\smallskip}
$32768$........... & $20686$ & $0.00$ & $1.00$ & $0.00$ & $10.90$  \\
$65536$........... & $41678$ & $0.00$ & $1.00$ & $0.00$ & $18.20$  \\
$131072$......... & $84050$ & $0.00$ & $1.00$ & $0.00$ & $30.85$  \\
\noalign{\smallskip}
\multicolumn{6}{c}{\tiny{- FAMILY 3: BINARY FRACTION -}} \\
\noalign{\smallskip}
$32768$........... & $20686$ & $0.00$ & $0.30$ & $0.10$ & $9.20$ \\
$65536$........... & $41589$ & $0.00$ & $0.30$ & $0.10$ & $15.30$\\
\noalign{\smallskip}
$32768$........... & $20872$ & $0.00$ & $0.30$ & $0.30$ & $8.45$\\
$65536$........... & $41760$ & $0.00$ & $0.30$ & $0.30$ & $13.75$\\
\noalign{\smallskip}
\hline 
\end{tabular} 
\end{table}

The initial density profile is given by a \cite{king_1966} model with a central concentration $W_0 = 7$ and an initial half-mass radius of $r_h = 1 $pc. This is preferable to a tidally limited cluster where the tidal radius from the King model is set equal to the Jacobi radius of the tidal field, which has proven to be unrealistic for modeling GC-like densities, since the resulting initial half-mass radii are on the order of 4-5 pc. This has the effect that the clusters are too extended at the beginning of their evolution and undergo core collapse only at the very end of their evolution. Furthermore, after expansion due to mass loss through stellar evolution, these clusters attain half-mass radii of 8-10 pc, much larger than what is observed in GCs today. For this reason we fixed the initial half-mass radius to 1 pc. This is more time-consuming because the smaller crossing times increase of the numbers of time steps.

All models started with a \cite{kroupa_2001} initial mass function (IMF, $\xi(m)\,{\rm d}m \sim m^{\alpha}\,{\rm d}m$) that has a Salpeter-like power law slope of $\alpha = -2.3$ for stars more massive than $0.5 M_{\odot}$  and a slope of $\alpha = -1.3$ for stars between $0.08$ and $0.5 M_{\odot}$. For the lower and upper mass limit we chose $0.1$ and $100 \, M_{\odot}$, respectively. This configuration leads to an initial mean mass of $\langle m \rangle = 0.63 M_{\odot}$. Primordial binaries were included for some models (see Section \ref{subsec:fam}) by randomly sampling the desired fraction of stars in the cluster to binaries with a thermal eccentricity distribution ($f(e) = 2e$) and a flat period distribution between 1 and $10^6$ days. The median ratio of the binary binding energy to the mean particle kinetic energy for these particular initial conditions is $e_{bin}/e_{kin}=2-4$ and decreases with increasing particle number.

Our clusters were set on a circular orbit around an external galaxy that follows the potential of an isothermal sphere ($\Phi(r) \propto \ln r$). The distance to the Galactic center was chosen to be $8.5$ kpc and the circular velocity of the galaxy V$_G$ was set to 220 \kms for all our simulations because we study Galactic GCs only. The particle integration was performed in an accelerated, but non-rotating reference frame where the cluster remained in the center and the galaxy moved around the cluster on a circular orbit. The forces generated by the Galactic potential and the stellar gravity were both applied to each star when it had advanced  on its orbit through the cluster.

Table \ref{tab:start} lists the initial parameters for our set of models, which from left to  right are the particle number ($N$), IMBH-mass fraction ($M_{\bullet}/M_{\rm{0}}$), black-hole retention fraction ($f_{\rm{ret}}$), binary fraction ($f_{\rm{bin}}$), and dissolution time ($t_{\rm{diss}}$, defined as the time the cluster needs to lose 95\% of its initial mass, see Section \ref{sec:time}). Due to an error in our mass-randomizing routine, some of the models with the same particle number contain the exact same mass. However, since every model starts with different initial parameters, this does not affect the individuality of each model. 
 
\subsection{Reference models} \label{sec:ref}

In Figure \ref{fig:ref} we depict the evolution of our reference models with N = 32k, 64k, and 128k stars. Shown are the bound mass fraction (a), the mean slope of the mass function $\alpha$ in the mass interval $0.3 \msun <  m < 0.8 \msun$ (b), the Lagrangian radii $r_{10\%}$ (c) and  $r_{50\%}$ (d), which include $10~\%$ and $50~\%$ of the cluster mass, respectively (the latter is the half-mass radius $r_h$), the remnant mass fraction (e), and the number of stellar-mass black holes (f) as a function of time.  

The evolution of the bound mass fraction in Figure \ref{fig:ref}.a shows a similar behavior for all models. After a steep drop from 1.0 to 0.7 in the first Gyr, which results mainly from the mass loss due to stellar evolution, the mass fraction drops almost linearly until the last few Gyr of the cluster lifetime. After this, the mass loss slows down. Apart from the increasing lifetime with particle number we did not find significant differences among these models in terms of their mass evolution. 

The next quantity we show is the evolution of the mass function slope $\alpha$. This is achieved by binning the mass intervals logarithmically, going from $m = 0.2 \ \msun$ to $10 \ \msun$ in 20 bins. We determined the slope of the mass function by fitting a power law to the points with $0.3 \msun < m < 0.8 \msun$, because this is the same fitting area as used for the observational data. The result is shown in Figure \ref{fig:ref}.b. As expected, the slope of the mass function increases with the age of the cluster because the cluster loses low-mass stars through two-body relaxation. As with the bound mass fraction, there are no significant differences between models with different particle numbers other than the time scaling (when these values are plotted as a function of the fraction of the total lifetime $t/t_{diss}$ all three models agree). 
   
Plots \ref{fig:ref}.c and d depict the evolution of the two Lagrangian radii that include 10~\% and 50~\% of the cluster mass, respectively. The core collapse is indicated by a drop in the 10~\% Lagrangian radius followed by an immediate rise due to binary heating. Surprisingly, the exact occurrence of that event does not seem to depend  on the particle number. While the 32k model undergoes core collapse first, the 128k core collapse still occurs before the 64k model reaches this point. We assume the reason for the absence of a clear trend for the occurrence of core collapse with particle number arises from stochastic effects caused by the low number of massive stars, which mainly drive the dynamical evolution of the cluster even for the 128k models.

The last two plots (Figure \ref{fig:ref}.e and f) show the evolution of the stellar remnants in the clusters. In Figure \ref{fig:ref}.e the remnant mass fraction increases with time. This is because low-mass stars are ejected more efficiently than high-mass stars. Consequently, the fraction of high-mass stars to low-mass stars constantly increases. The steep rise at the beginning is due to initially rapid stellar evolution and the consequent formation of stellar remnants. The evolution of the stellar-mass black hole numbers in the clusters in Figure \ref{fig:ref}.f shows a roughly  exponential behavior. For a model with 128k stars, the number of black holes already drops to half of the initial value within 1 Gyr. After 5 Gyr all black holes have been ejected from the cluster. To lead the eye, we applied rough logarithmic fits to the evolution of black-hole numbers (dashed lines). This agrees well with previous results \citep[e.g.][]{sigurdsson_1993,kulkarni_1993,miller_2002,oleary_2006}, which show that stellar mass-black holes eject themselves from the cluster core by building a dynamically decoupled subcore through mass segregation \citep{spitzer_1969}. A recent study by \citet{sippel_2013} has shown, however, that it is possible to retain a large number of stellar-mass black holes in the cluster after a Hubble time with even small retention fractions of $10\%$. The initial half-mass radius in their simulations was a factor of 6 larger than our values. This leads to longer crossing and relaxation times and consequently to a delayed dynamical evolution of the cluster. This is shown by the fact that the models of \citet{sippel_2013} do not undergo core collapse after 10 Gyr. Therefore, the ejection rate of the stellar-mass black holes is also low. We note, however, that models with initial half-mass radii of several parsec end up with physical sizes too large compared with those of the majority of galactic GCs after their dynamical evolution. A comparison of panel c) and f) indicates that the core collapse in all models occurs after the cluster has lost its stellar-mass black holes.

   \begin{figure}
  \centering
   \includegraphics[width=0.49\textwidth]{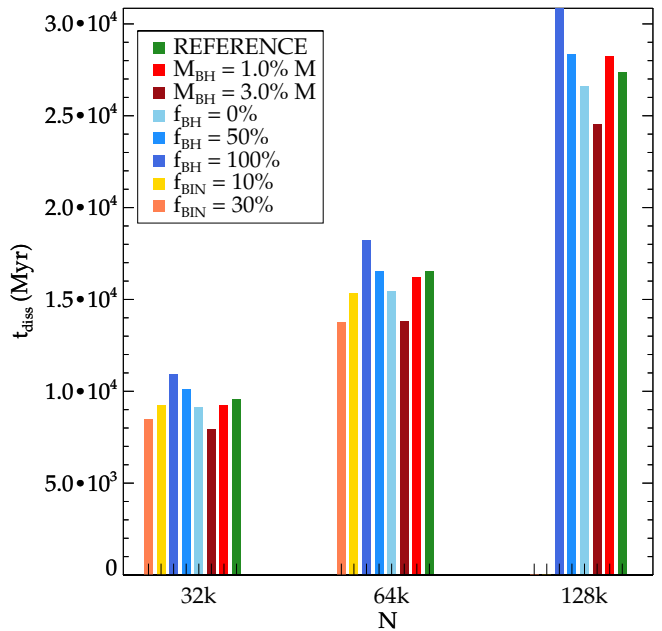}
      \caption{Lifetimes of all clusters as a function of their initial particle number.}
         \label{fig:tdiss}
   \end{figure}

\subsection{Model families} \label{subsec:fam}

We ran models with different IMBH masses, black-hole retention fractions, and binary fractions, which we sorted to three different families. To use realistic values for the IMBH masses we considered several scaling relations such as the $M_{\bullet} - M_{\rm{bulge}}$ relation \citep{magorrian_1998, haering_2004} and the $M_{\bullet} - \sigma$ relation \citep{ferrarese_2000,gebhardt_2000,gultekin_2009}. When these empirical correlations are extrapolated, they predict IMBH masses of $M_{\bullet} \sim 0.001 - 0.03 \ M_{\rm{tot}}$ in GCs \citep{Goswami_2012}. This corresponds to black-hole masses  $M_{\bullet} \sim 10^2 - 5 \times 10^3 \ M_{\odot}$  in an average GC with a mass of $M_{\rm{tot}} = 1.5 \times 10^5 \ M_{\odot}$. These masses were found by \cite{portegies_2004}, who used $N$-body simulations to model the runaway merging of massive stars in the dense star cluster MGG-11. Their values, however, should be treated as upper limits, since stellar mass loss of the 'runaway' star was not included \citep[see]{glebbeek_2009}. Taking into account these predictions, we used two different black-hole masses: $1\%$ and $3\%$ of the total cluster mass. For models with black-hole masses $< 1\%$ the actual black-hole mass is $< 800 \ M_{\odot}$ even in simulations of N = 128k stars and the central black hole has a high chance of being ejected by encounters with massive stars.
   
   \begin{figure*}
  \centering
   \includegraphics[width=0.33\textwidth]{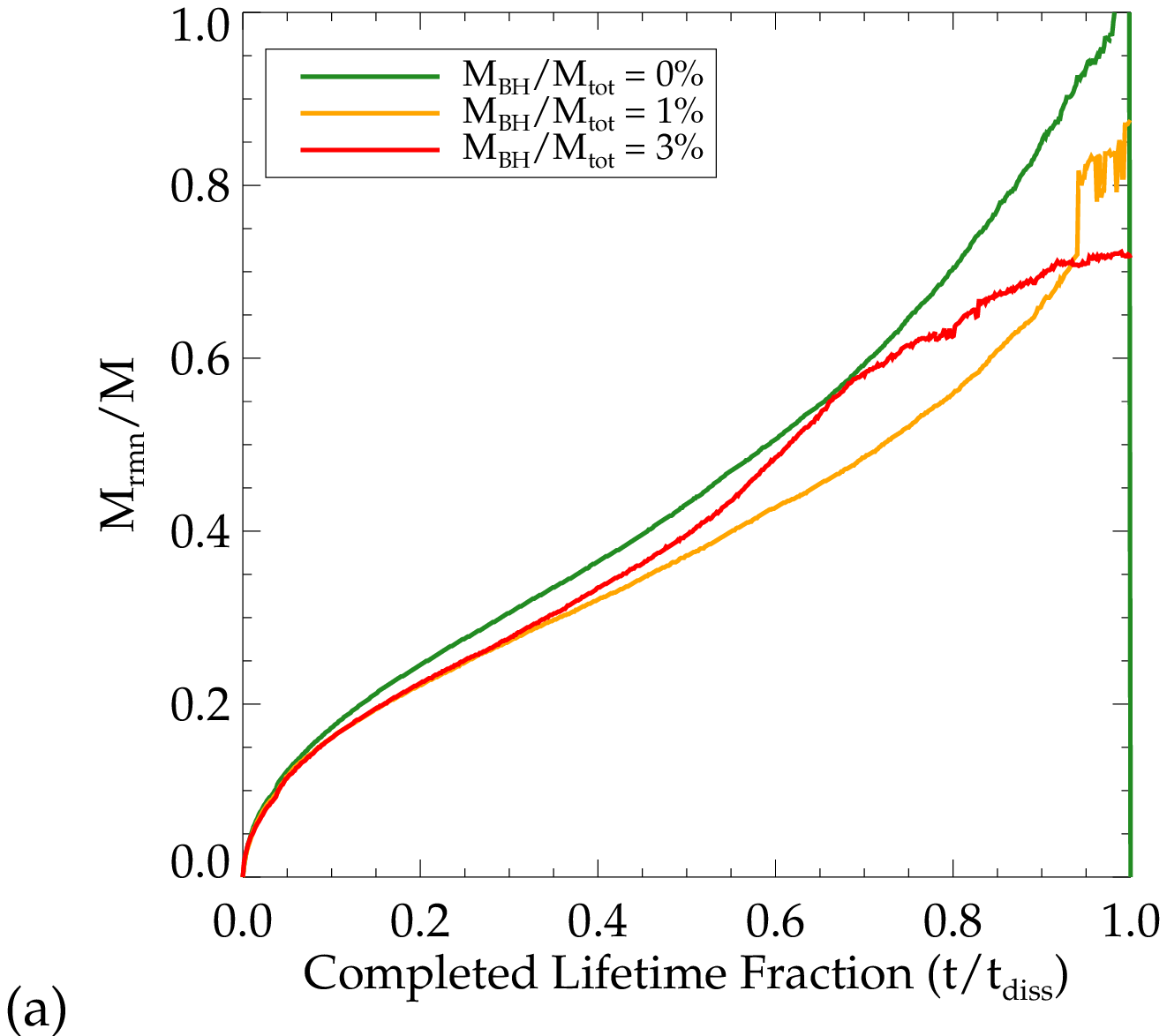}
      \includegraphics[width=0.33\textwidth]{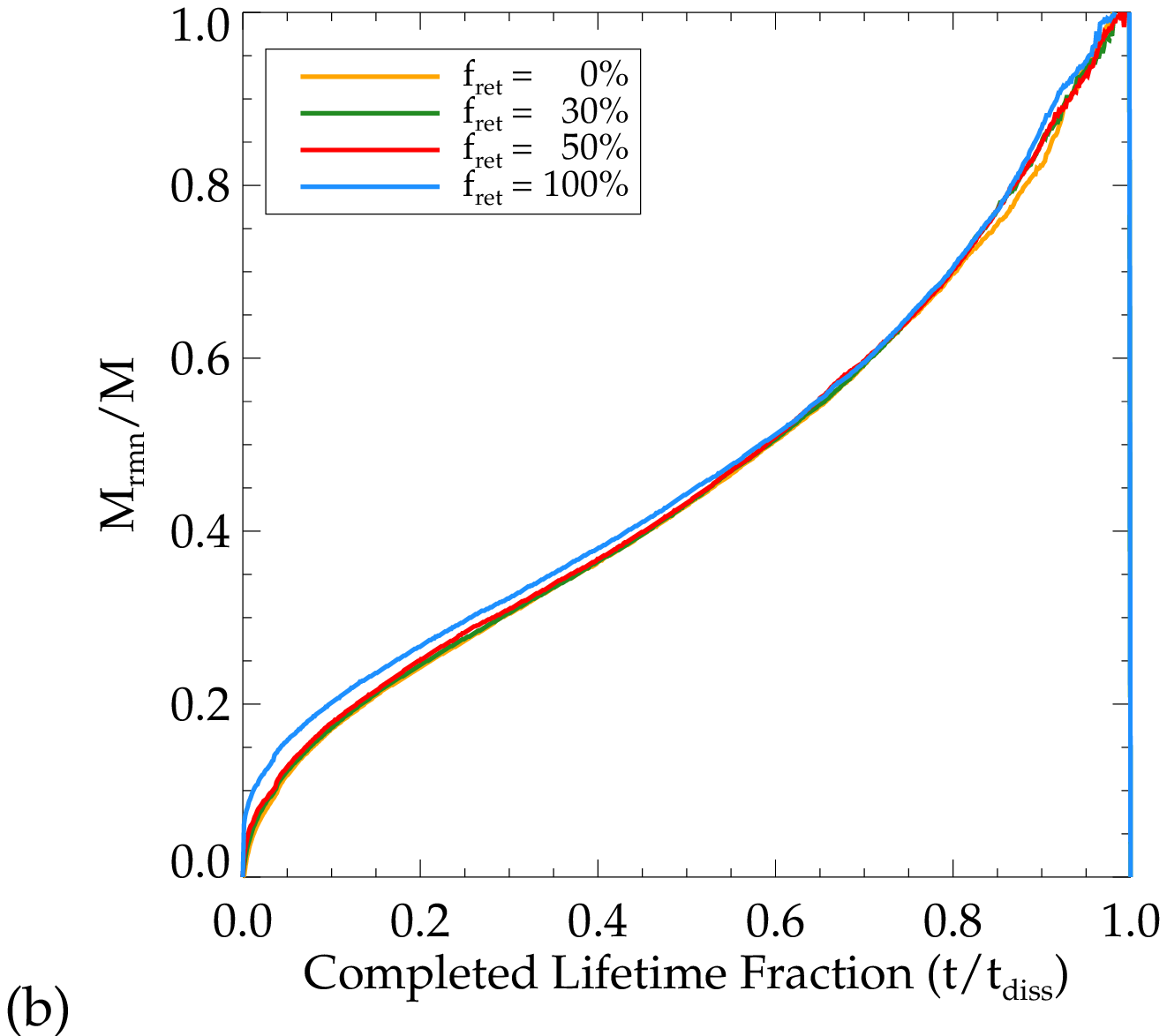}
      \includegraphics[width=0.33\textwidth]{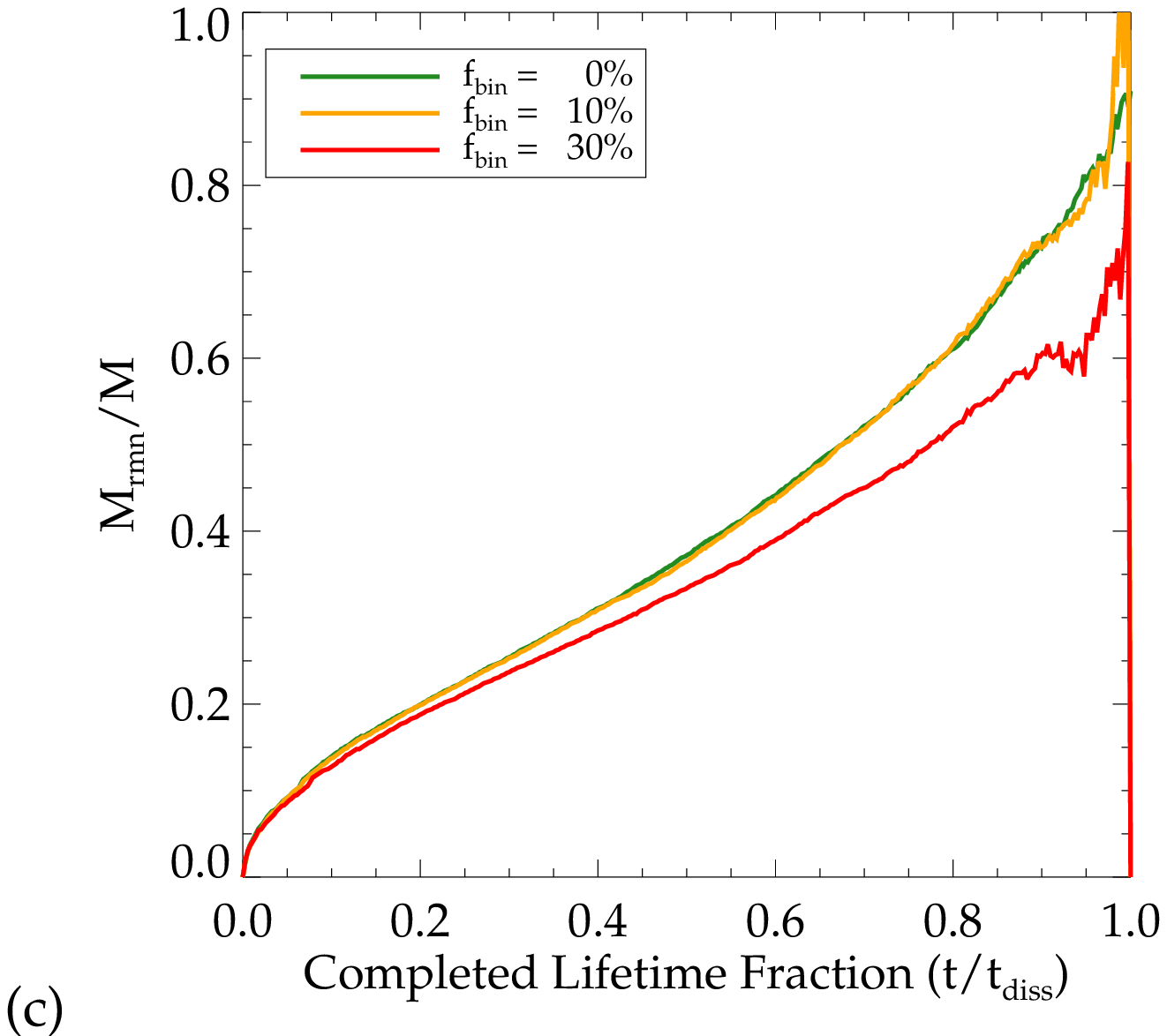}
      \caption{Remnant mass fractions as a function of the completed lifetime fraction for simulations with different a) IMBH masses, b) black-hole retention fractions, and c) primordial binary fractions. All models contain 128k particles, except for the binary-fraction models, which are limited to 64k stars.}
         \label{fig:remn}
   \end{figure*}   
   
The retention fraction gives the fraction of stellar remnants that remain in the cluster after their formation. It is assumed that very compact objects such as black holes and neutron stars suffer a kick velocity during their formation that is often higher than the escape velocity of the cluster \citep{lyne_1994}. Theoretical estimates for the retention fraction of neutron stars range from 5~\% to 20~\% \citep{drukier_1996}. For black holes, however, the range of possible values is not well-constrained since the physics of the formation of black holes is still not fully understood.  For this reason, we assumed a constant retention fraction of $10~\%$ for neutron stars and varied the retention fractions of the black holes from 0~\% (no black hole remains in the cluster) to 100~\% (all black holes that are formed remain in the cluster). For white dwarfs we always assumed a retention fraction of $100 \%$ since they do not suffer any kick velocities \citep{davis_2008}. 

Binaries are important for the dynamical evolution of GCs and need to be considered in our models. To test how primordial binaries affect the models, we introduced a third family, in which we varied the fraction of binaries in the initial conditions of the model from 10~\% to 30~\%. This covers the current typical binary fraction in GCs, which is about 10~\% \cite[e.g.][]{rubenstein_1997,davis_2008,dalessandro_2011}. Since calculating the individual time steps of the binary systems is very time-consuming, we limited this exercise to a maximum particle number of 64k for the models with primordial binary fractions.


\section{Observables}\label{sec:obs}

The main goal of this study is finding observables in GCs that give information about the presence of a possible intermediate-mass black hole at their centers. In this section we study the effects of IMBHs on their environment to pin down important tracers for central black holes in GCs. We especially concentrate on mimicking observations, that is, measuring cluster properties in the same way as observers would do to compare our results with observational studies.

\subsection{Lifetimes} \label{sec:time}

The lifetime, or dissolution time $t_{\rm{diss}}$, gives information about the dynamical evolution of the model as a function of its internal properties. In this work, we define the dissolution time as the time when the cluster has lost 95\% of its initial mass. 

In Figure \ref{fig:tdiss} we compare the lifetimes of our model families with the reference models (green bars). As expected, there is a clear trend with particle number and slight differences (of about $\sim 10 -20 \%$) between models with different initial conditions. The higher the mass of the central IMBH, the shorter the lifetime of the cluster. The opposite trend is observed for models with different black-hole retention fractions. The model with the longest lifetime is the model with a black-hole retention fraction of 100 \%. However, it is remarkable that the difference in lifetimes is only about $20\%$ and therefore does not show a strong dependence on whether the cluster hosts an IMBH, a large number of stellar-mass black holes, or primordial binaries.


The primordial binary fraction weakly affects the lifetime of the cluster model. Encounters with binaries can result in flybys, exchanges, ionizations, and sometimes mergers \citep[see][]{luetzgendorf_2012b}. Some of these encounters result in high-velocity stars that exceed the escape velocity of the cluster. This can lead to a higher ejection rate than in clusters without tight binaries. 

We compared our results with the lifetimes found in \citet{baumgardt_2003}, who studied $N$-body simulations of GCs in tidal fields with different galactocentric distances and eccentricities. With a neutron star retention fraction of 100\% and no stellar-mass black holes, the initial conditions assumed by these authors slightly differ from ours. However, since their stellar masses are cut off at $15 \msun$, the effective retention fraction of neutron stars is lower and the models from their family II are comparable with our $f_{\rm ret} = 0\%$ models (see Table \ref{tab:start}). Except for the 32k model, the lifetimes of the models in this study are $\sim 10 \%$ longer than those found in of \citet{baumgardt_2003}. These discrepancies are most likely caused by the underfilling initial conditions of our simulations in contrast to the tidally filling clusters of \citet{baumgardt_2003}.

\subsection{Remnant fractions}

Another quantity we considered is the fraction of the cluster mass constituted by stellar remnants. As remnants we counted all stars within one tidal radius classified as either a white dwarf, a neutron star, or a black hole. The results are displayed in Figure \ref{fig:remn}, where we compare the remnant fractions for models with different black-hole masses, retention fractions, and primordial binary fractions as a function of the completed lifetime fraction (i.e. $t/t_{diss}$). The remnant fractions increase monotonically. This occurs because of the increasing number of remnants as a result of stellar evolution and also because low-mass stars are ejected first. For all models, this increase continues until the end of the cluster's lifetime. For models with non-zero IMBH masses and primordial binary fractions, however, the remnant fraction shows a slower increase than that of the reference model. For models with a high fraction of stellar-mass black holes, the remnant fraction does not exhibit a strong variation. An explanation for the lower remnant fraction in models with IMBHs and primordial binaries could be that the lower degree of mass segregation caused by the IMBH distributes massive stars and remnants farther outside in the cluster where they can be more easily ejected. For binaries, the effect might be caused by the higher survival chance of low-mass stars in binaries. Owing to their higher total mass, they remain in the cluster longer than to single low-mass stars, and the remnant fraction increases more slowly. 

In summary, remnants are ejected more efficiently from a cluster with a central massive black hole, while no notable effect is found for clusters with different black-hole retention fractions. This is in contrast with the result found by \cite{kruijssen_2009}, who used semi-analytic models to study the evolution of the stellar mass function in star clusters. These authors proposed that the remnant fraction could also decrease toward the end of a cluster's lifetime if a large portion of stellar-mass black holes were retained.

   \begin{figure*}
  \centering
   \includegraphics[width=0.33\textwidth]{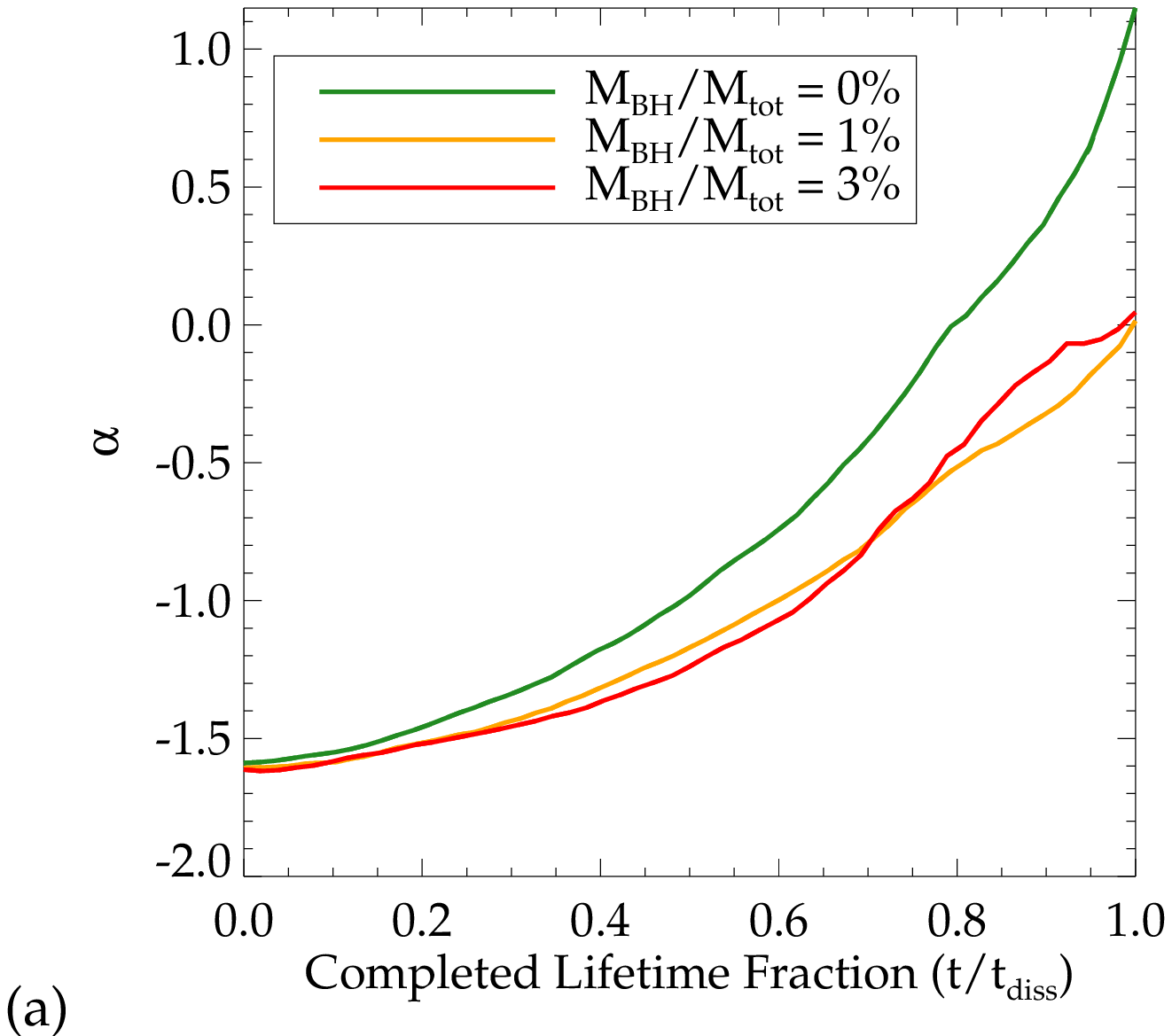}
      \includegraphics[width=0.33\textwidth]{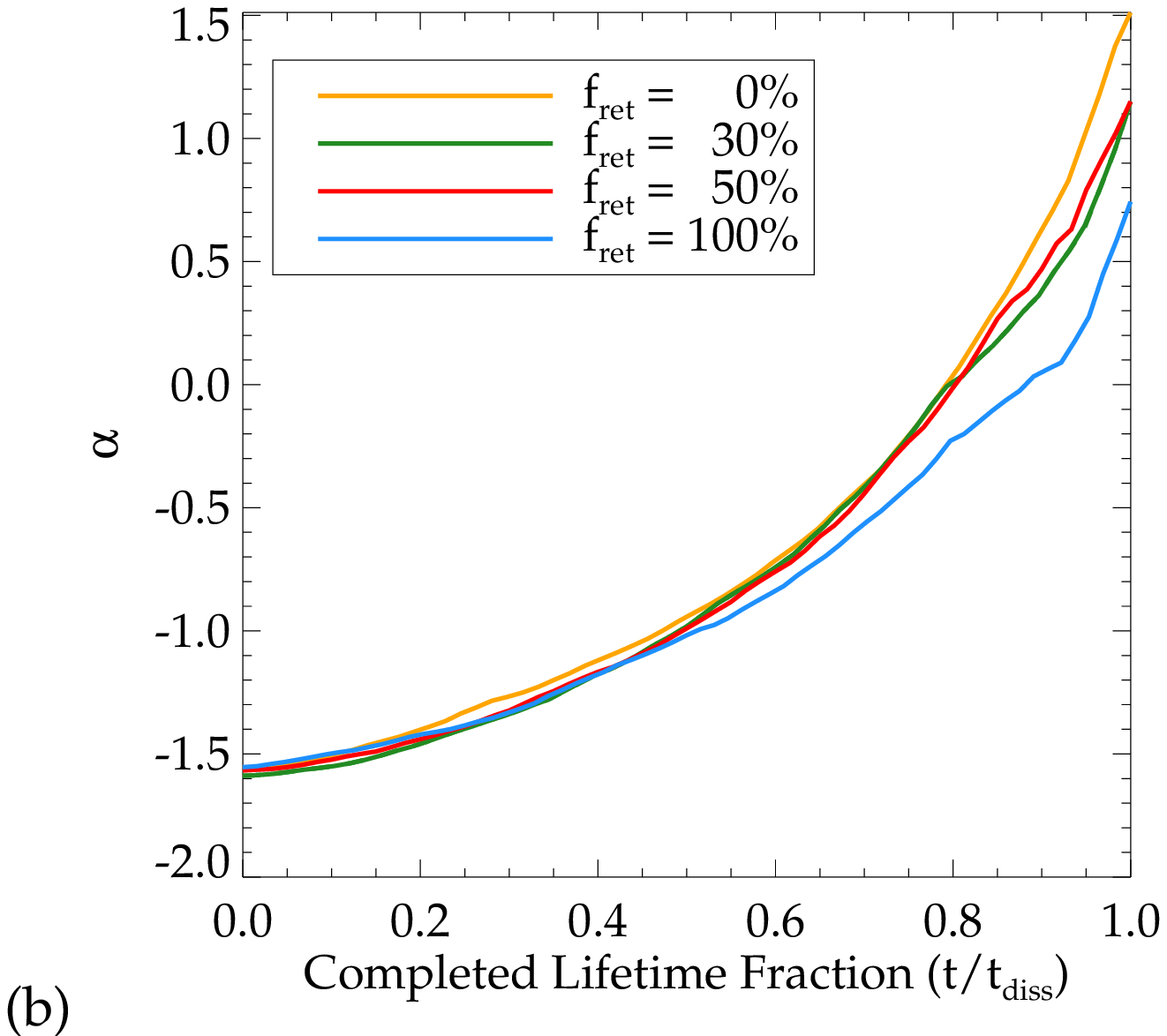}
      \includegraphics[width=0.33\textwidth]{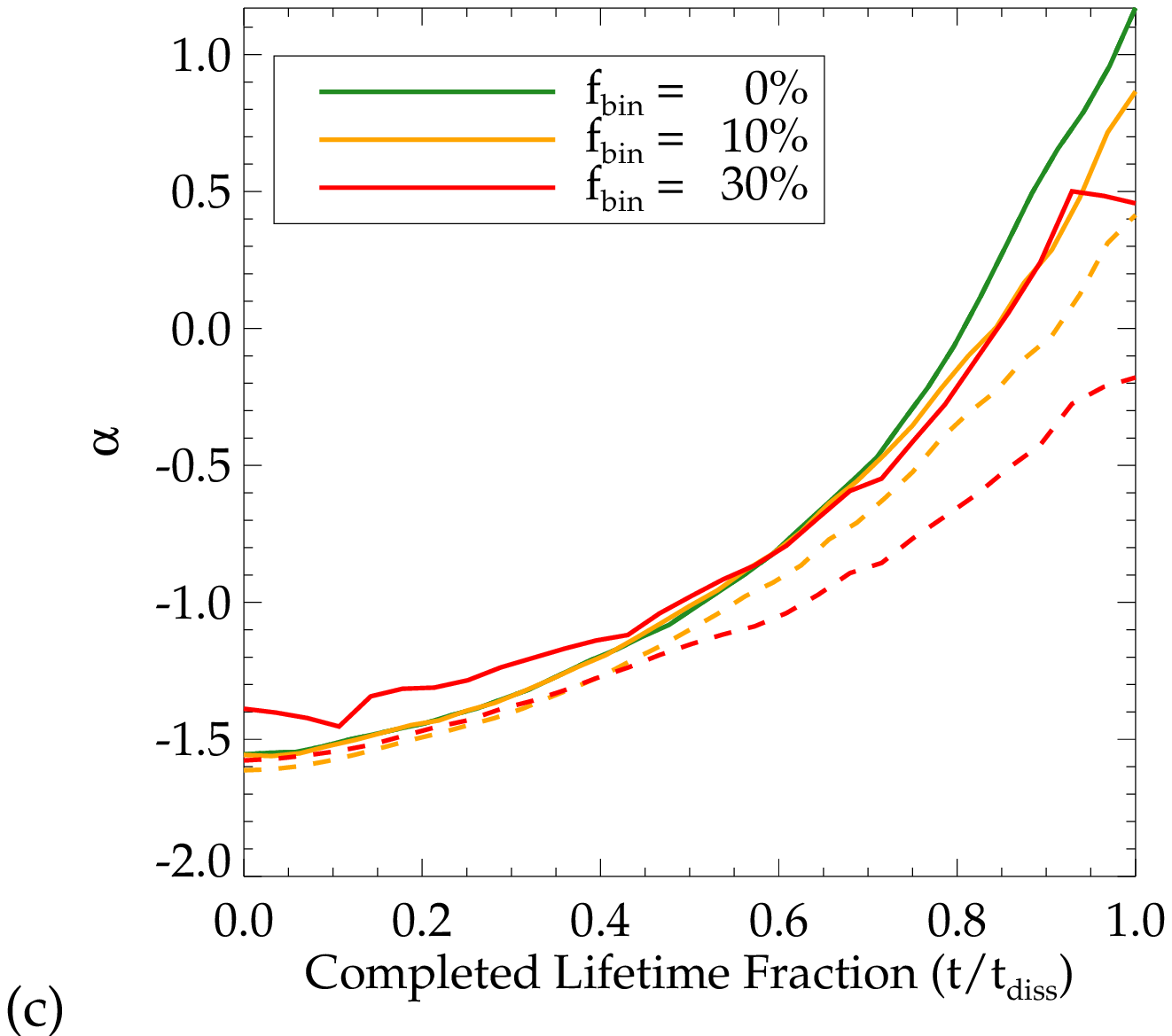}
      \caption{Mass-function slopes (for $0.3 \msun < m < 0.8 \msun$) as a function of the completed lifetime fraction for simulations with different a) IMBH masses, b) black-hole retention fractions, and c) primordial binary fractions. All models contain 128k particles, except for the binary-fraction models, which are limited to 64k stars. The dashed lines indicate the true mass function slope, while the solid lines indicate the mass function slopes as measured by an observer, that is, counting binaries as single stars.}
         \label{fig:alpha}
   \end{figure*}

\subsection{Mass function slope}

The slope of the mass function is a measure for the evolutionary state (or dynamical age) of a GC because it gives information about how many stars and which types of stars were ejected. We used the method described in Section \ref{sec:ref} to derive the mass function slopes in the mass range of $0.3 \msun < m < 0.8 \msun$ for our models. We stress that our goal is to compare the derived properties of the simulations with observations. For binaries, we therefore took the total luminosity of the system and transferred this into mass using the existing $M(L)$ relation of single stars in the cluster. In this way we  ensured that the binaries contribute to the mass function as they would in observations where the binary components cannot be separated.  In Figure \ref{fig:alpha} we plot the slope of the mass function as a function of the completed lifetime fraction (solid lines). For comparison we overplot the 'true' evolution of $\alpha$ (dashed lines) as it would be measured if all binary stars could be resolved.  As expected, low-mass stars gain high velocities due to two-body relaxation and escape the cluster. Therefore, the slope of the mass function at the low-mass end changes from negative to positive and increases monotonically. This behavior is observed for all our simulations. However, the details of the change in the slope differ from model to model.   

As shown in panel a) of Figure \ref{fig:alpha}, the IMBH mass has a significant influence on the mass function slope. This most probably arises from the fact that a central black hole ejects high-mass stars more efficiently than the non-IMBH model, as already demonstrated in the previous section. In panel b) of Figure \ref{fig:alpha} the evolution of $\alpha$ seems to be very similar irrespective of how many stellar-mass black holes are retained by the cluster. Only for a retention fraction of 100\% a slight difference in $\alpha$ becomes noticeable. 

Binaries also cause a slower rise of the mass function slope and therefore a lower depletion in low-mass stars. As shown in panel c) of Figure \ref{fig:alpha}, the simulation that starts with a primordial binary fraction of $30 \%$ in the end has the lowest \textit{true} value of $\alpha$ of all simulations (dashed lines). As described in the previous section, this can be explained by the lower ejection rate of low-mass stars that reside in binaries. However, this is only measurable when all binaries are resolved. A large portion of binaries can bias the observed value of the mass function slope to higher values (solid lines). The differences increase with cluster lifetime and range from $\Delta \alpha = 0.05 - 0.5$ and $\Delta \alpha = 0.2 - 0.7$ for the $10\%$ and the $30\%$ binary fraction model, respectively. 


   \begin{figure*}
  \centering
   \includegraphics[width=0.33\textwidth]{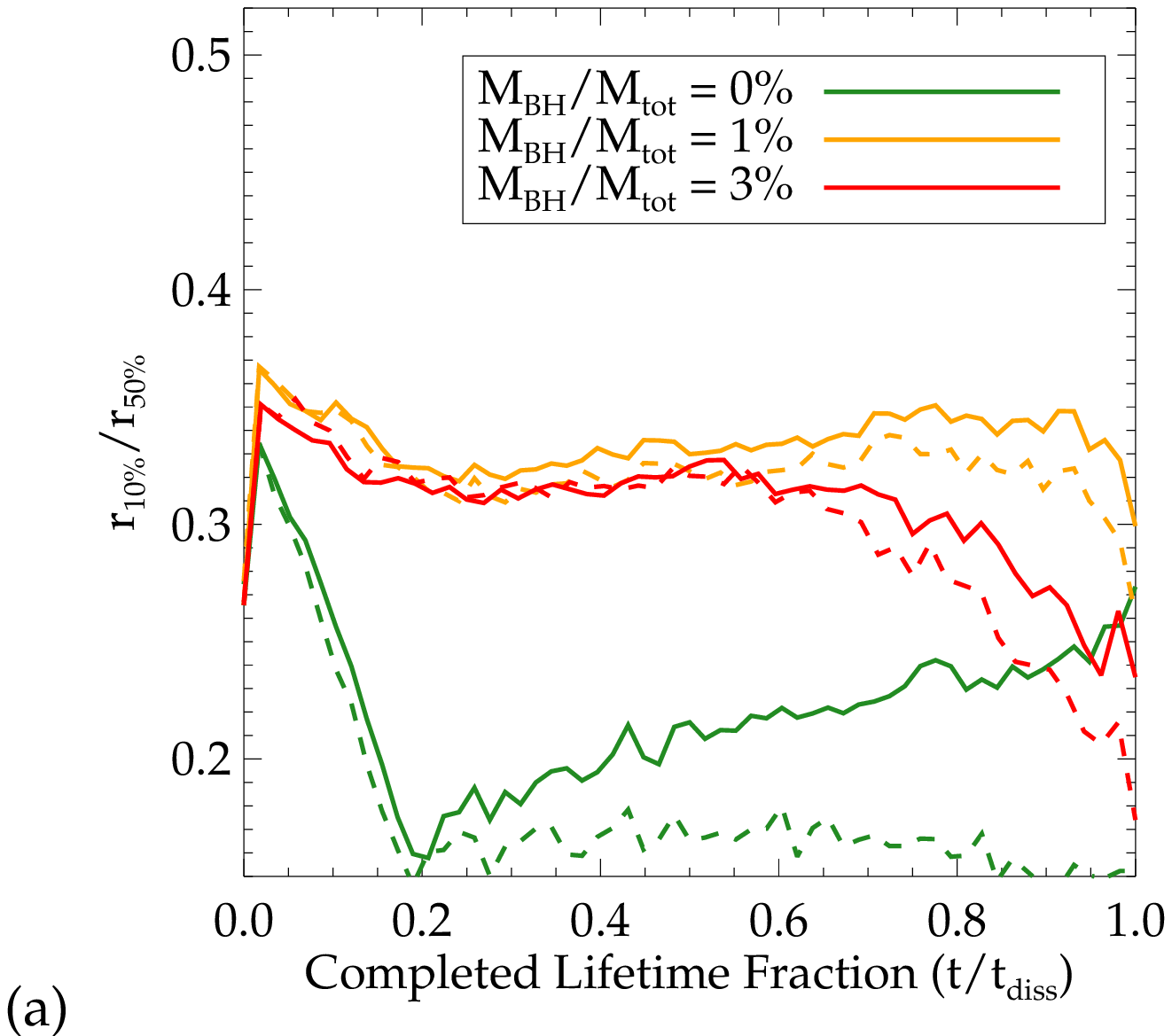}
      \includegraphics[width=0.33\textwidth]{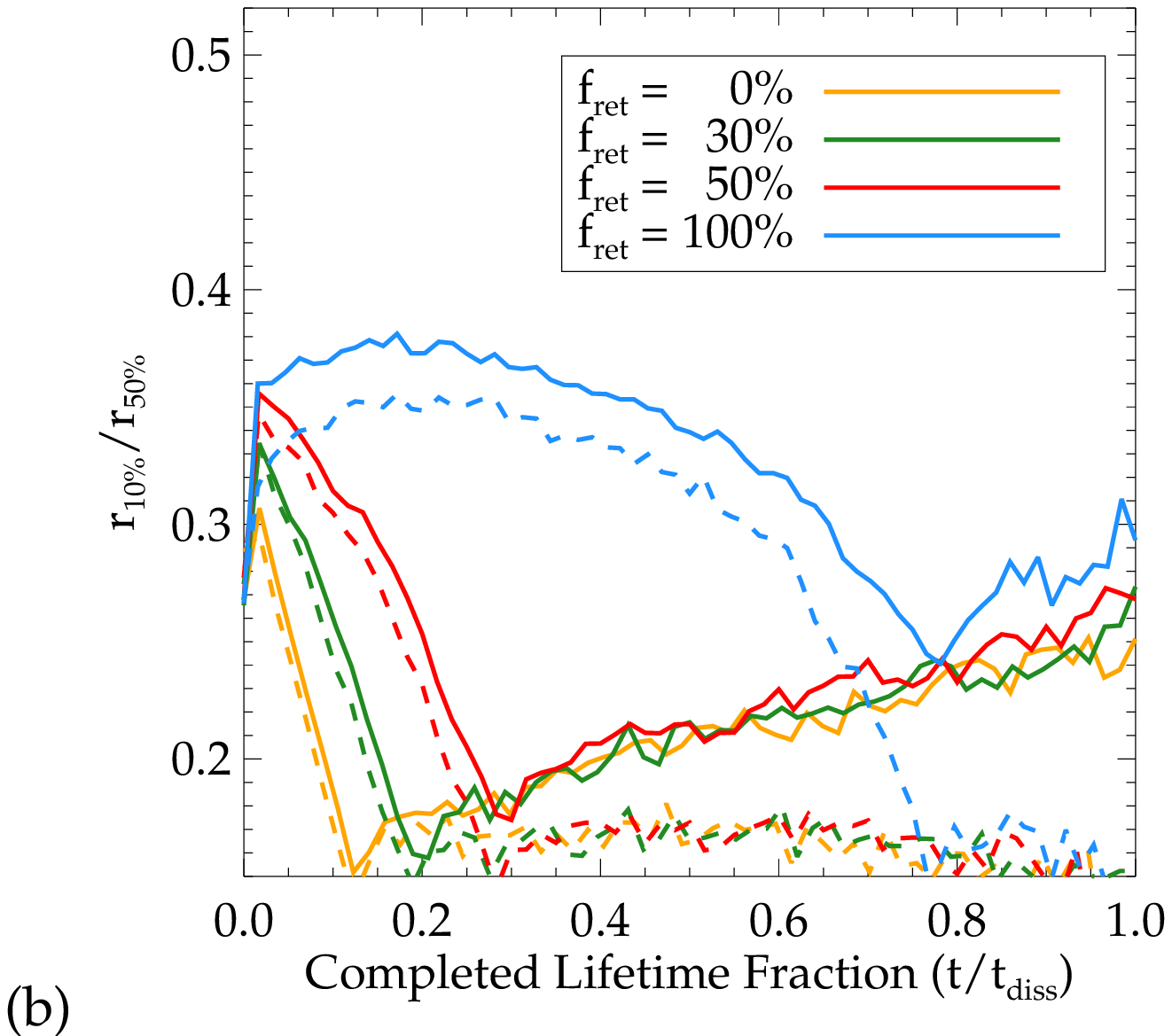}
      \includegraphics[width=0.33\textwidth]{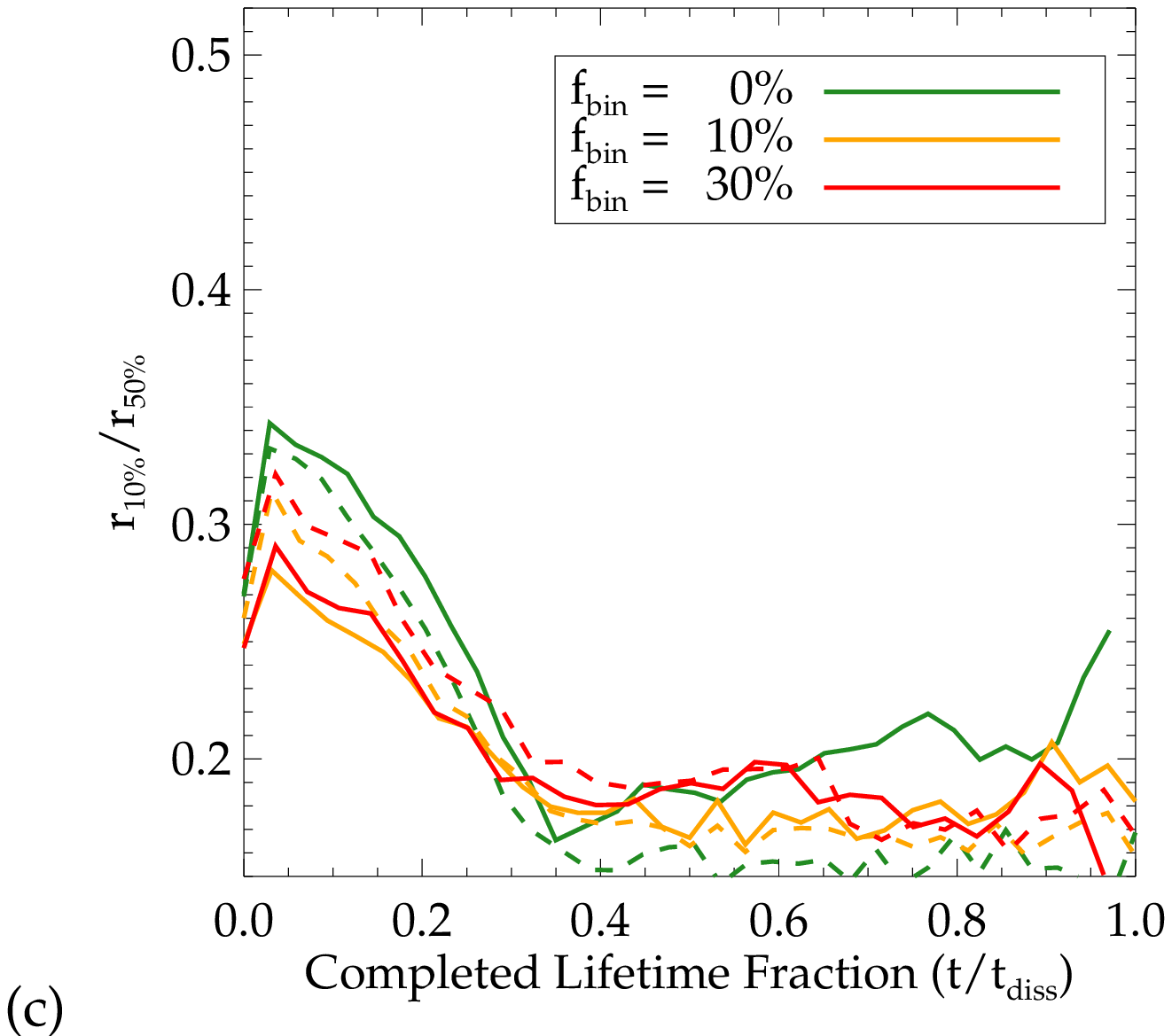}
      \caption{Characteristic radii ratios ($r_{10\%}/r_{50\%}$, i.e., deprojected radii containing $10\%$ and $50\%$ of the stars) as a function of the completed lifetime fraction for simulations with different a) IMBH masses, b) black-hole retention fractions, and c) primordial binary fractions. All models contain 128k particles, except for the binary-fraction models, which are limited to 64k stars. The dashed lines indicate the $r_{L,10\%}/r_{L,50\%}$ (deprojected radii containing $10\%$ and $50\%$ of the mass) evolution when taking the Lagrangian radii.}
         \label{fig:radii}
   \end{figure*}
   
\subsection{Structural properties} \label{subsec:radii}

The final quantities of interest are the cluster's structural properties, such as the characteristic radii containing 10\% and 50\% of its stars. Again, we aim to compare these quantities with observations and therefore derive the values in the same way as we do for the observed clusters. To obtain these observables in a non-parametric and consistent way, we applied several steps: First, we derived a projected surface brightness profile for stars brighter than $M_v = 8.5$ mag in 15 logarithmic bins from $r_{min} = 1''$ (by placing all our clusters at a distance of 10 kpc, this corresponds to $\sim 0.05$ pc) out to the tidal radius of the cluster. In each bin the stellar number density is computed for three  orthogonal projected directions. For the final density we took the average of the three projected directions and used the scatter as the error. With this method we derived a projected density profile similar to that observed for a real cluster.  As a next step we deprojected the stellar density profile of each cluster using the projected profiles and a multi-Gaussian expansion method \citep{emsellem_1994} to compare this with observations. This parametrization has the advantage that a smooth deprojected profile can be easily obtained since the deprojection of any Gaussian again is a Gaussian. By integrating the deprojected profiles, we obtain radii that include different percentages of the total number of stars, such as $r_{50\%}$ and $r_{10\%}$.

In Figure \ref{fig:radii} we show the evolution of the ratio $r_{10\%}/r_{50\%}$ with time. One prominent feature of this plot is the strong dip in the ratio at about $t/t_{\rm diss} = 0.2$ ($\sim 5$ Gyr) that indicates core collapse. The plots demonstrate that for the standard model with a conservative black-hole retention fraction of 30~\% and no primordial binaries or IMBH, core collapse is a very prominent feature in the evolution of the structural parameters (see Figure \ref{fig:radii}.a, green line). The presence of an intermediate-mass black hole at the center of the cluster entirely prevents core collapse and $r_{10\%}/r_{50\%}$ is nearly constant over the cluster's lifetime (Figure \ref{fig:radii}.a). A similar effect arises if a large number of stellar-mass black holes is present. As shown in Figure \ref{fig:radii}.b, the model with a retention fraction of 100~\% undergoes core expansion instead of core collapse, which is delayed until the final stages of its evolution. This underlines the dependence of the time of the core collapse on the black-hole retention fraction. Surprisingly, the binary fraction does not seem to have a strong effect on the concentration of the cluster. Core collapse occurs at the same time as for the model without any primordial binaries, and the overall evolution of  $r_{10\%}/r_{50\%}$ is very similar. However, this result should be treated with care because of the the low-number statistics of the 64k models. 

To estimate the degree of mass segregation in our models, we furthermore considered the evolution of the Lagrangian radii $r_{L,10\%}$ and $r_{L,50\%}$. These quantities indicate the radii that contain $10\%$ and $50\%$ of the total cluster mass (i.e., also including remnants), respectively. Because of two-body relaxation and mass segregation, remnants sink to the center of the cluster, causing a higher concentration of mass. This causes the ratio of $r_{L,10\%}$ and $r_{L,50\%}$ to be lower than $r_{10\%}$ and $r_{50\%}$ that were derived from the stellar distribution alone. Therefore, studying the difference between the two ratios gives an estimate of the degree of mass segregation. In Figure \ref{fig:radii} the evolution of $r_{L,10\%}/r_{L,50\%}$ is indicated by the dashed lines. It is notable that the ratios of the characteristic radii differ for all our models after some time of cluster evolution. This indicates that mass segregation is present in all the clusters, but its degree differs from model to model. Especially after core collapse, the remnant segregation is much higher than observed in the visible stars for clusters without IMBHs or low stellar-mass black hole retention fractions. The lowest mass segregation is observed for models with non-zero IMBH masses. This result is of great interest for understanding the connection between observed quantities and the actual dynamical stage of the globular cluster.

\section{Comparison with observed GCs}\label{sec:comp}

This section describes the comparison of globular cluster $N$-body simulations with observed properties. This may help to explain empirical correlations and identify new cluster candidates for hosting IMBHs.

\subsection{$\alpha - c$ plane}

\cite{demarchi_2007} and \cite{paust_2010} studied the mass function slopes $\alpha$ as a function of the concentration parameter $c$ for a total sample of 36 Galactic GCs. As mentioned in the previous section, the slope of the mass function is a measure of the dynamical age of the cluster. The higher the slope (i.e. higher value of $\alpha$), the more low-mass stars were already lost through two-body relaxation in the external tidal field. Therefore, one would expect a direct correlation between this slope and other quantities that are correlated with the dynamical time scale of the cluster (see Figure \ref{fig:alpha}). One of these quantities is the concentration $c$, which is a measure for the central density of the cluster. The more advanced the cluster evolution, the closer it is to core collapse, and in turn, the higher is its central concentration. The observations, however, contradict this picture by showing the opposite relation between mass-function slope and concentration \footnote{Note that the authors measured the mass function slope for the mass range $0.3 \msun < m < 0.8 \msun$. This includes the \textit{break} of the mass function that is usually located at $0.5 \msun$. For this reason, the measurements are very sensitive to uncertainties in total magnitudes and distances and should therefore be treated with care.}.

To measure the concentration parameter $c$ that we can compare with the observations, we computed a projected density profile (as described in Section \ref{subsec:radii}) and fit an isotropic \citet{king_1966} model to the simulation data. This allowed us to derive the structural parameters (e.g. $c$, $r_c$, and $r_h$) from the best fit. 

  \begin{figure*}
  \centering
   \includegraphics[width=0.45\textwidth]{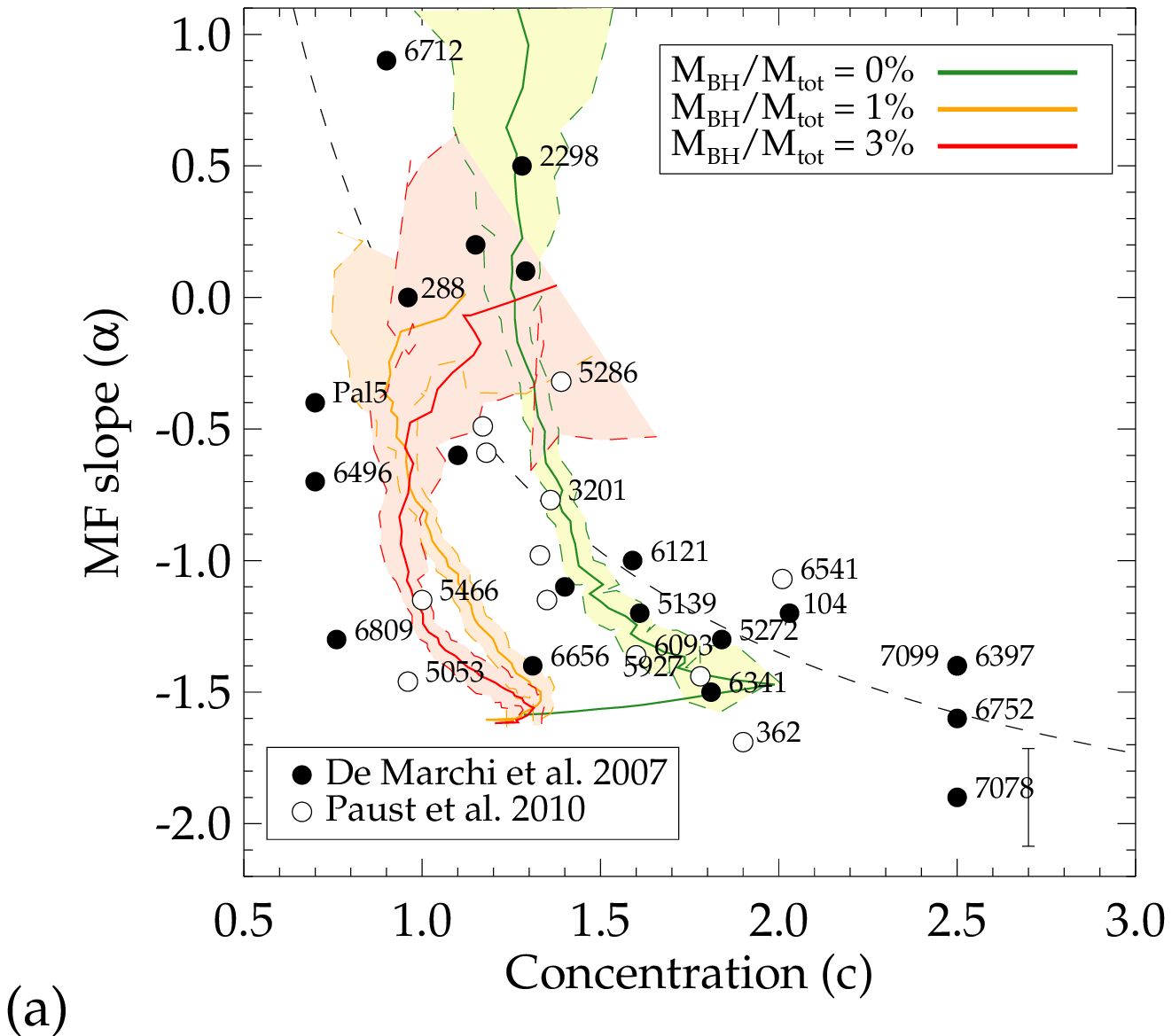}
            \includegraphics[width=0.45\textwidth]{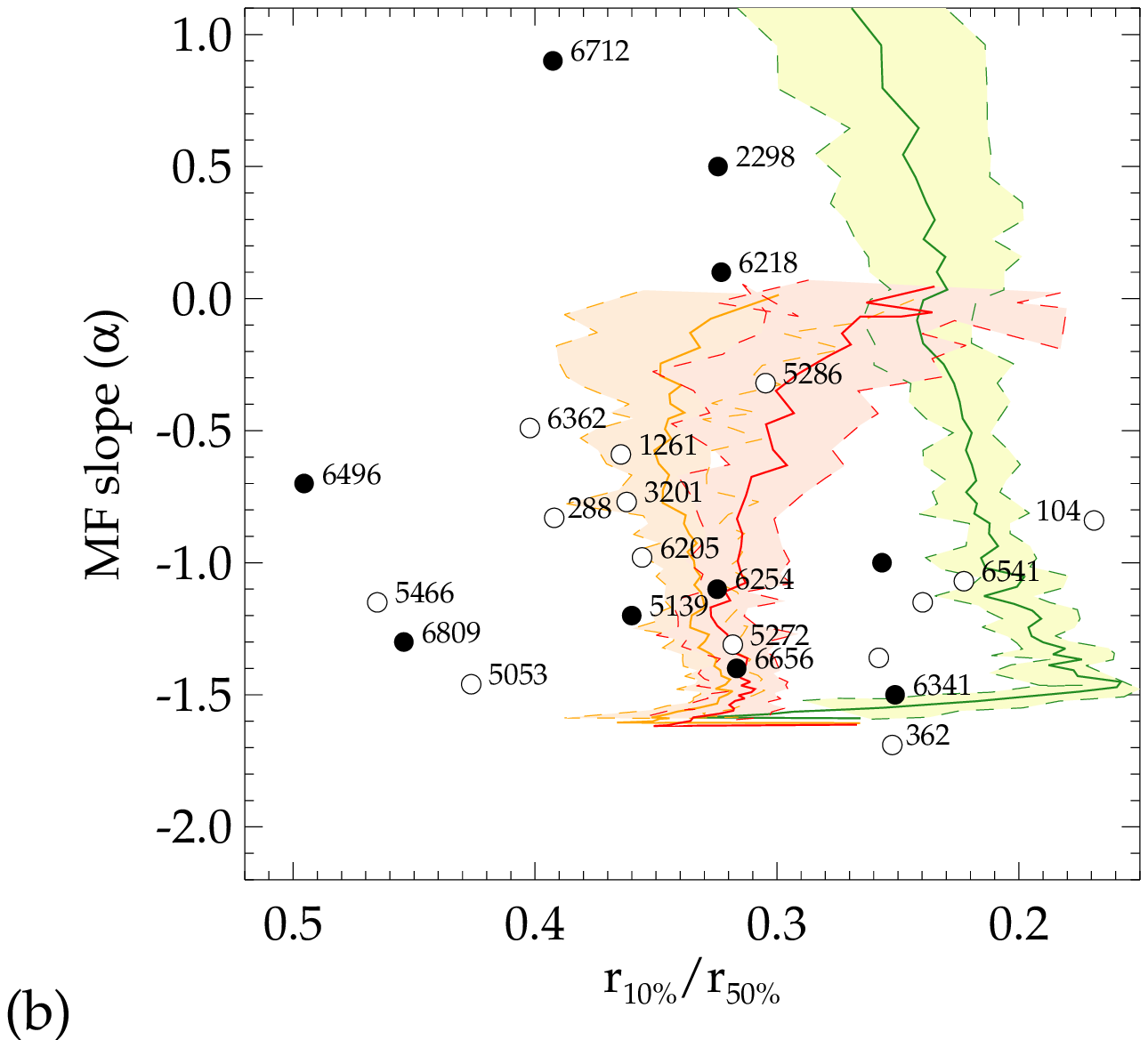}
      \includegraphics[width=0.45\textwidth]{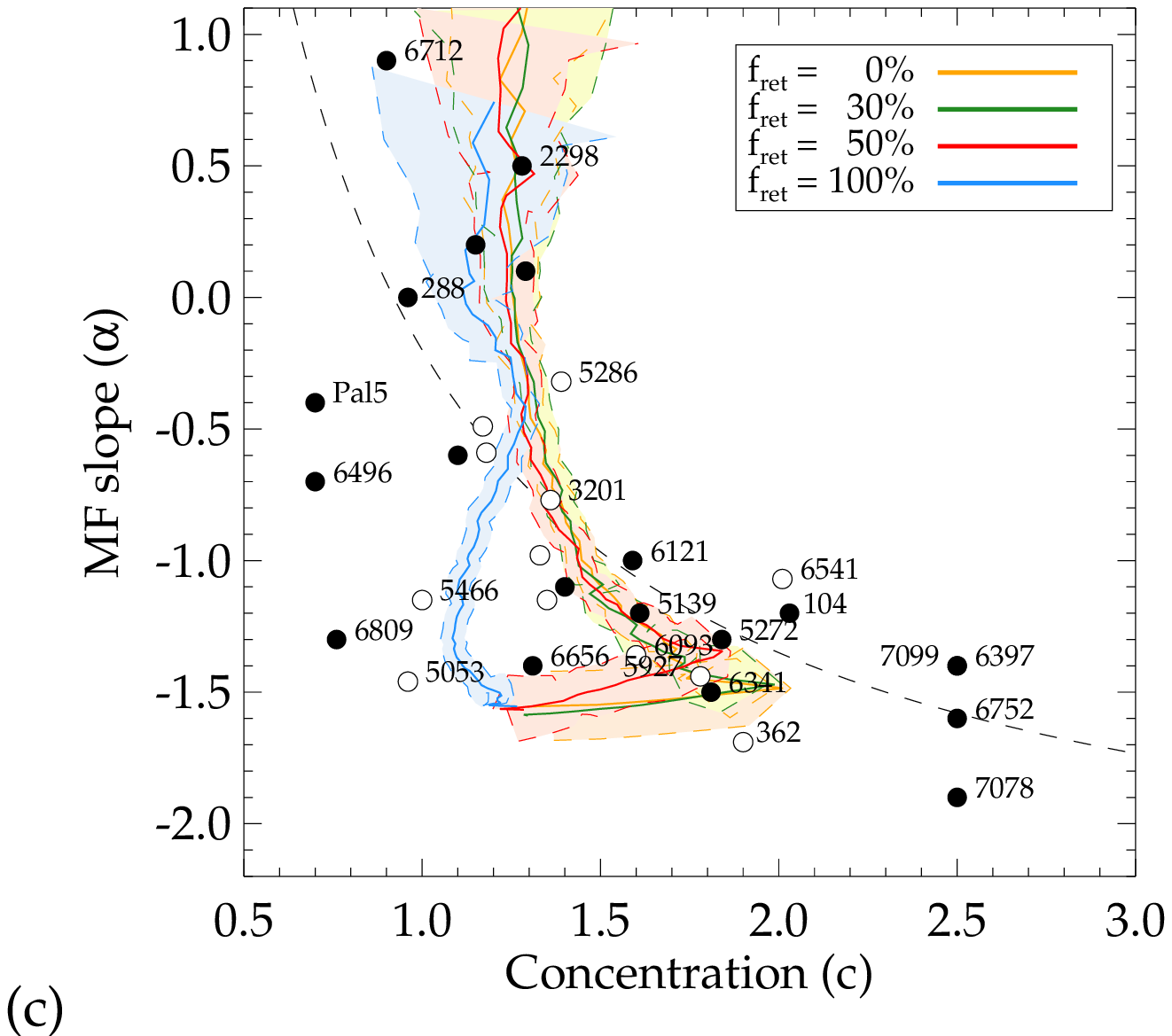}
            \includegraphics[width=0.45\textwidth]{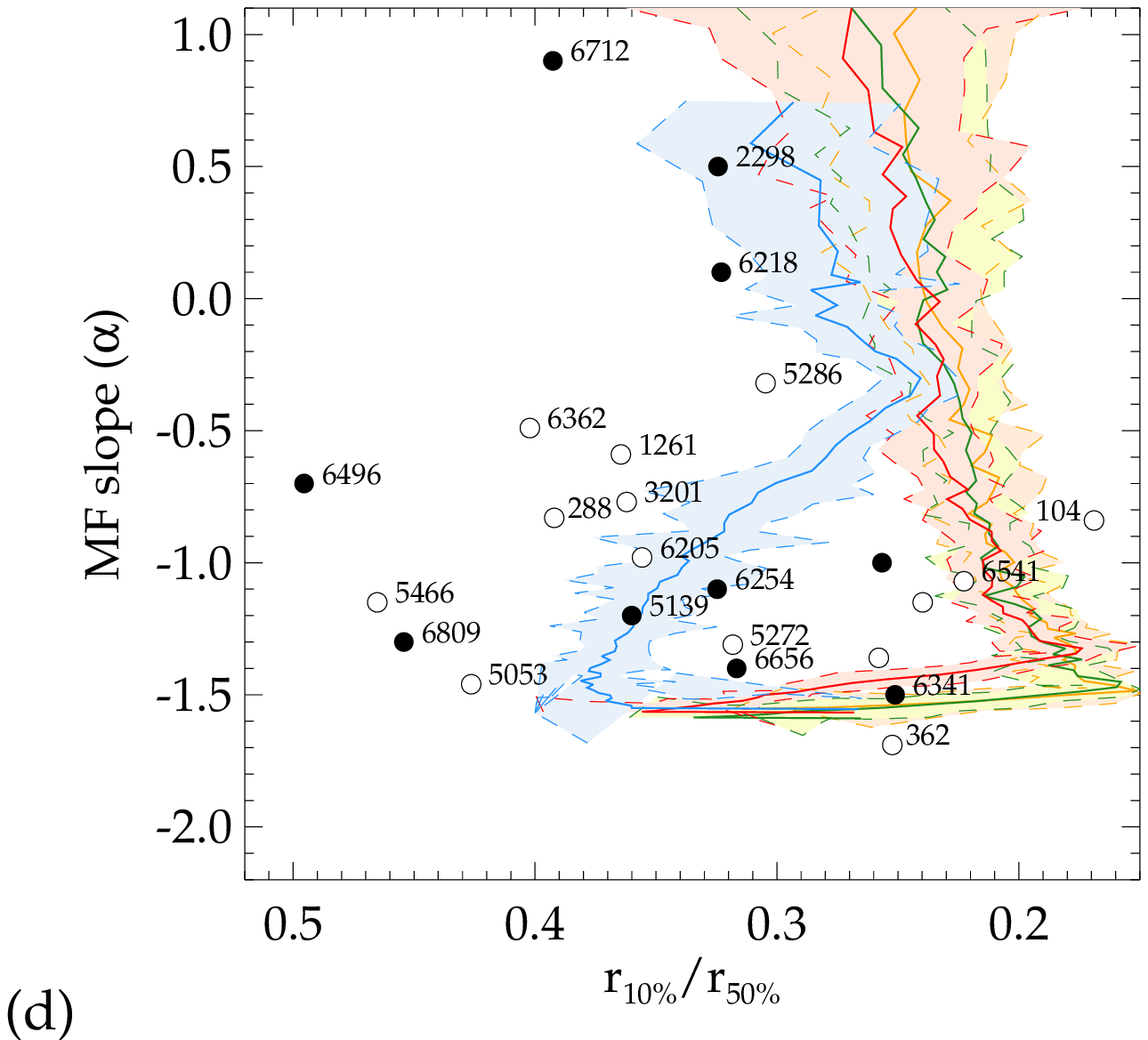}
      \includegraphics[width=0.45\textwidth]{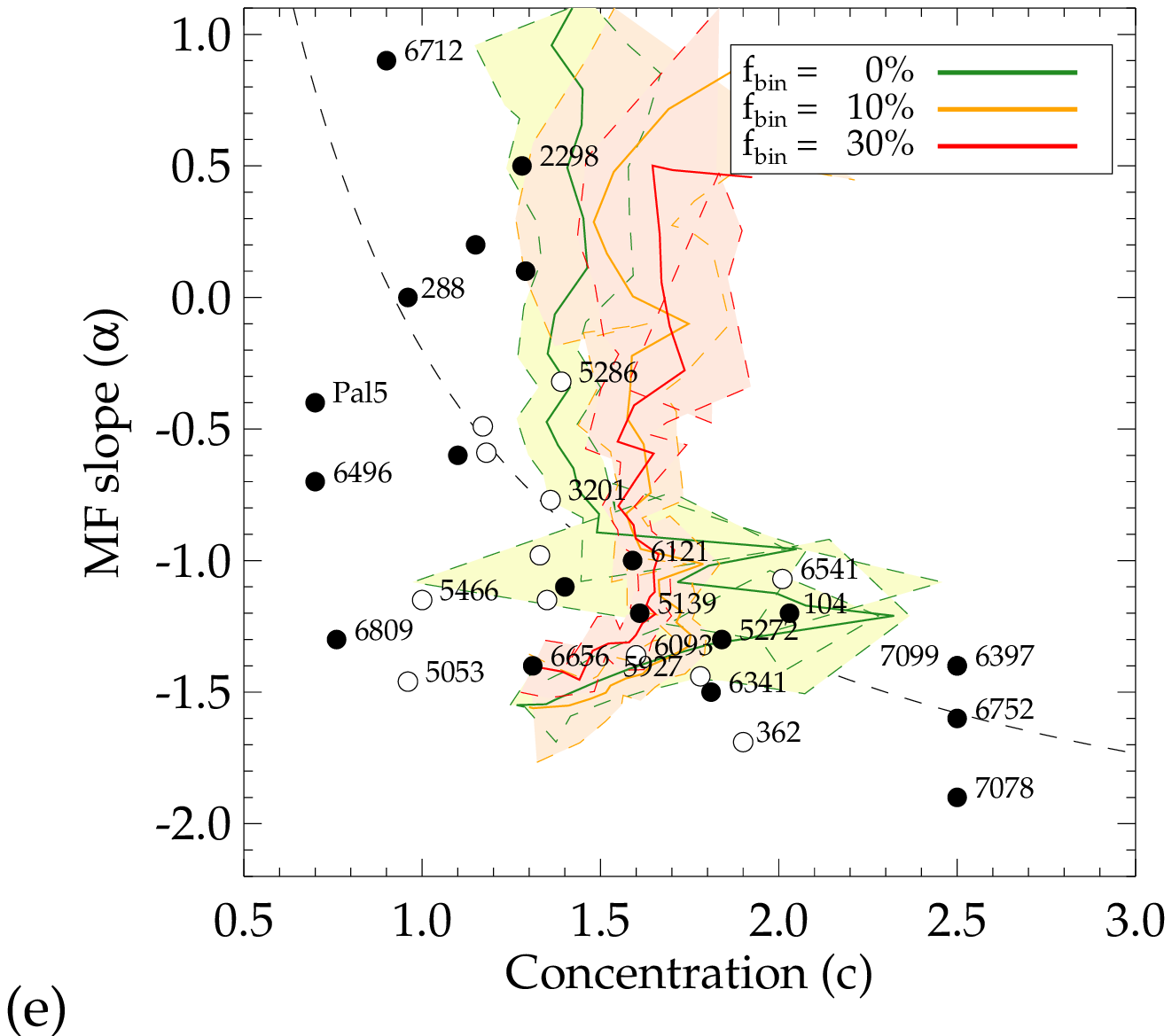}
       \includegraphics[width=0.45\textwidth]{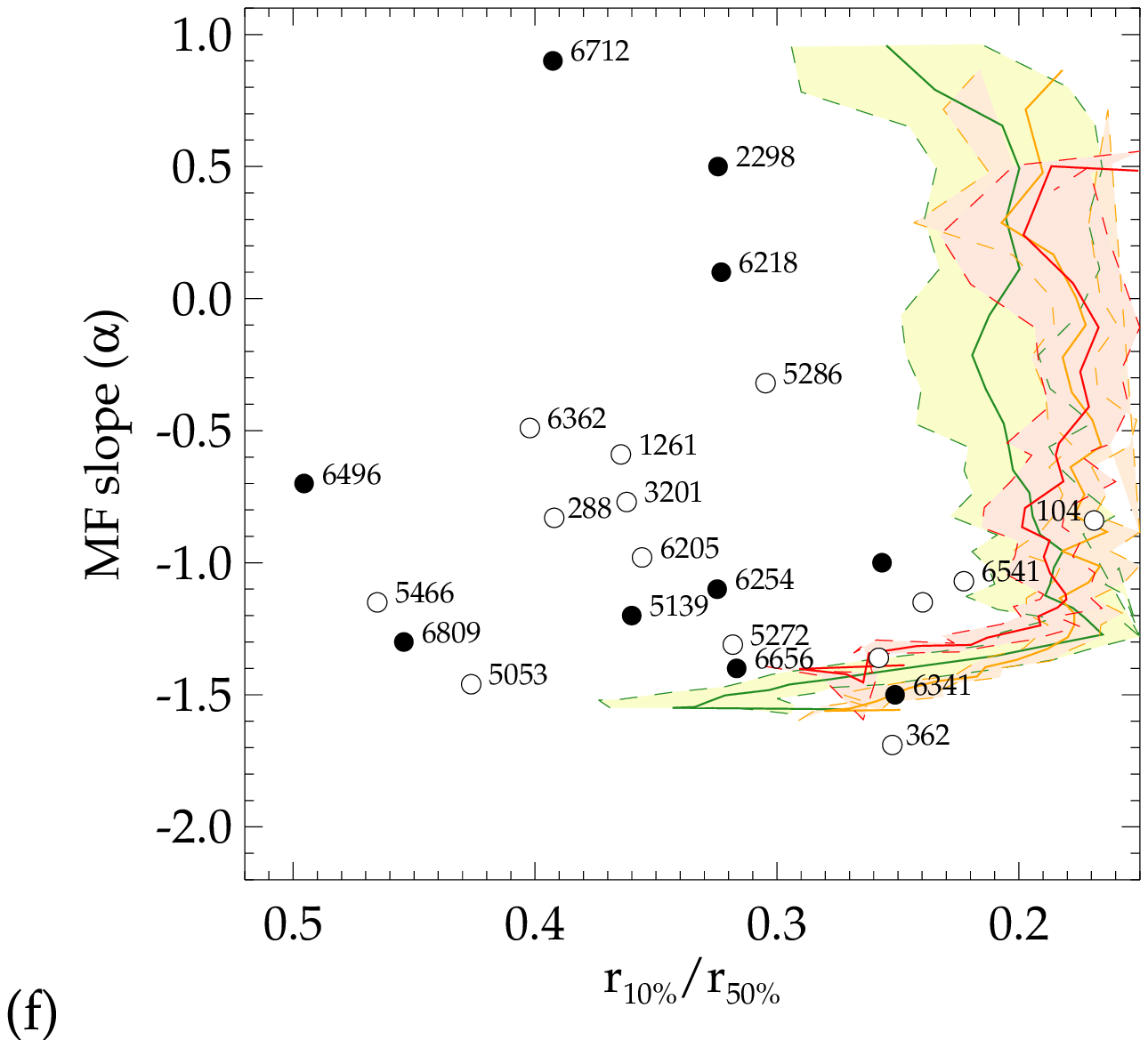}
      \caption{$\alpha$ - c plane (left panels) and $\alpha - r_{10\%}/r_{50\%}$ plane (right panels) of observed GCs compared with $N$-body simulations. Shaded areas mark the $3 \sigma$ limit of the simulations. Clusters falling within these limits and not overlapping with other models are considered to be well-represented by these models. The error bar in panel a) is representative for the uncertainties in $\alpha$ for the observed clusters. The labeled clusters are discussed in Section \ref{sec:con}.}
         \label{fig:ac}
   \end{figure*}

To compare with the work of \cite{demarchi_2007} and \cite{paust_2010}, we derived $\alpha$ (for $0.3 \msun < m < 0.8 \msun$) and $c$ at points in time spaced by $500$ Myr from the cluster models. Figure \ref{fig:ac} (left panels) displays the tracks on top of the observed data of \citet[][filled circles]{demarchi_2007} and \citet[][open circles]{paust_2010}. As before, we compare models with different IMBH masses, black-hole retention fractions and primordial binary fractions. The figure shows that our standard model (without an intermediate-mass black hole and retention fraction of 30~\%) undergoes core collapse after several Gyr and  moves back to lower concentrations during its evolution (green line). Furthermore, the figure shows that the cluster without a black hole behaves in the expected way and $c$ increases with $\alpha$ before it undergoes core collapse. After core collapse, however, the dependence of these two values is reversed because the cluster core slowly expands again due to binary heating. Comparing the green line with the data, it seems plausible that clusters with medium-to-high concentrations and relatively unevolved mass functions can be represented by a no-IMBH model with a medium retention fraction. Clusters with central massive black holes are presented by the yellow and the red lines. These models do not undergo core collapse and represent the data points with low concentrations very well. A similar agreement was found by \cite{trenti_2010} for a single IMBH run.

The suppression of core collapse can also be produced by many stellar-mass black holes, as demonstrated by the model with the 100~\% black-hole retention fraction (Figure \ref{fig:ac}.c, blue line). The high concentration of black holes in the center also enlarges the core of the visible stars and prevents core collapse because it functions as an extra heating source. As already observed for the structural properties, binaries do not affect the concentration of the cluster and the tracks are similar for binary and non-binary models. 

\subsection{$\alpha - r_c/r_h$ plane}

Another and preferable way of comparing these data with the models is using the derived non-parametric structural parameters $r_{10\%}$ and $r_{50\%}$ instead of the concentration parameter. The ratio of these radii is a more reliable indicator  than the ratio provided by the parametric approach, because a King model does not always provide a good fit to the data after core collapse. 


To extract the characteristic radii from the observational data we took all clusters that are included in the sample of \cite{paust_2010} and \cite{demarchi_2007} and applied the same procedure as for the $N$-body data. We matched their clusters with the sample of \cite{trager_1995} and \cite{McLaughlin_2005} to obtain surface-brightness profiles and masses for as many clusters as possible. The overlap is substantial and we obtained a sample of 22 clusters. 
For each cluster we then deprojected the surface brightness profile using the smoothed profiles from \cite{trager_1995} and the multi-Gaussian expansion method described in Section \ref{subsec:radii}. We integrated the deprojected surface-brightness profiles to obtain the radii $r_{10\%}$ and $r_{50\%}$ and used the total masses obtained by \citet{McLaughlin_2005}.

From Figure \ref{fig:ac} (right panels) we conclude that the ratio of the two characteristic radii is a good replacement for the concentration parameter. The observed data follow a similar distribution as in the $\alpha - c$ plot (i.e., there appears to be a lack of clusters with low $r_{10\%}/r_{50\%}$ value and high $\alpha$). As already concluded from the $\alpha - c$ analysis, models with large cores and low values for the mass function slope are good candidates for hosting a central IMBH. However, the degeneracy with the models with a $100~\%$ black-hole retention fraction needs to be treated with caution.  The agreement between the observations and the different $N$-body runs can be exploited to identify candidate GCs for hosting an IMBH. We describe this in the next section.

%

\section{Summary and discussion}\label{sec:con}

We have investigated the effect of intermediate-mass black holes, stellar-mass black hole retention fractions, and primordial binary fractions on the properties of GCs evolving in a tidal field. We ran $N$-body simulations using the GPU-enabled version of NBODY6 with 32k, 64k, and 128k stars. We studied the effect of the different initial conditions on the cluster lifetime, remnant fraction, mass function, and structural parameters. In addition, we compared the results of the simulations with observational data from the literature and found good agreement. Owing to the specific shape of the King profile, the concentration parameter $c$ is a poor representation of the cluster's internal properties. Especially after core collapse, a King model is not able to reproduce the central cusp in any of our models and the concentration is systematically underestimated. For that reason we also computed the ratio of the radius containing $10\%$ and $50\%$ of the stars in the cluster $r_{10\%}$ and $r_{50\%}$, which is a more accurate quantity than the parametric King fit. 

When comparing the different models, we found that a cluster with central IMBH has a shorter lifetime than the rest of the models. This is caused by the enhanced ejection of stars due to lower mass segregation. On the other hand, models with high black-hole retention fractions live the longest of all simulations. A possible explanation for the extended lifetimes is that large numbers of stellar-mass black holes would expand the core and prevent core-collapse, thereby slowing down dynamical evolution. We stress that the difference in cluster lifetime is about $20\%$ and the effects of IMBHs, stellar-mass black holes, and binaries are therefore modest.

Our study also showed that a central IMBH in a GC causes the remnant fraction to increase more slowly than in a cluster without a central black hole. However, a similar effect is reached with a high ($30\%$) fraction of primordial binaries. Again, the reason for this behavior might be found in the lower degree of mass-segregation that causes high-mass stars and remnants to be distributed in the outskirts of the cluster where they can be easily ejected. For the evolution of the mass function in the simulations, we found that the central IMBH as well as a primordial binary fraction of $30\%$ lead to a delay of the depletion of low-mass stars because high-mass stars are ejected. Therefore, the mass function slope has a lower value (i.e. is less depleted in low-mass stars) at the end of the cluster's lifetime than for models without an IMBH or binaries. 

When comparing the models with the observations in the plane of the mass function slope versus the cluster's concentration, we found that the IMBH models clearly explain the observed data with low concentrations. By contrast, the models without an IMBH immediately undergo core collapse. Our more extensive parameter survey thus confirms the initial suggestion by \cite{trenti_2010}. The analysis also showed that the same effect can be achieved when considering a black-hole retention fraction of $100\%$. In this case the black holes in the cluster center survive long enough to prevent core collapse over a long timescale. However, it seems very unrealistic to retain all black holes in the cluster after formation. We therefore conclude that GCs that lie in the upper left side of Figure \ref{fig:ac}, that is, those with low concentrations but depleted mass functions, are good candidates for hosting (or having hosted) an IMBH at their center.

To find the best candidates for clusters hosting an IMBH we selected those that agreed within $3 \sigma$ (derived from the spread in the values of $\alpha$ and $c$ within 500 Myr) with one of the IMBH runs. Figure \ref{fig:ac} shows the comparison with the models. The color-shaded areas mark the $3 \sigma$ limits of the simulations. Only a few clusters can be identified as clear IMBH or non-IMBH candidates. The majority of the clusters lie in the white areas and are therefore not clearly defined. For the $\alpha - c$ plot, the best IMBH candidates are NGC~288, NGC~5466, and NGC~6656 because they directly overlap with the $3 \sigma$ regions of the IMBH simulations. But NGC~5053, NGC~6809, NGC~6496, and Pal~5 are also promising because their values are closer to those predicted from the IMBH models than the non-IMBH models. Furthermore, NGC~6712 could be considered if the lifetime of the IMBH models were extended to reach higher values for $\alpha$. NGC~5272, NGC~6341, NGC~5927, NGC~6093, and NGC~3201 agree with the non-IMBH model in their values of $\alpha$ and c. Analogous to the IMBH candidates, the high-concentration models to the right of the no-IMBH models, e.g., NGC~6397 or NGC~7078, can also be considered as clusters that most likely do not host an IMBH at their center. NGC~5286 agrees with all three models, which makes it challenging to compare with previous studies. \citet{feldmeier_2013} found the inner kinematics of NGC~5286 to be consistent with a $15~000 M_{\odot}$ black hole at its center. However, as mentioned earlier, the $\alpha - c$ plane for GCs might not be the best parameter to compare with our simulations because the King model is a poor representation of an evolved cluster's structural parameters. 

Comparing the non-parametric radii $r_{10\%}$ and $r_{50\%}$ in the same way for  we find that NGC~1261 , NGC~3201, NGC~5286, NGC~5272, NGC~6254, and NGC~6656 overlap with the $3 \sigma$ regions of the IMBH models. The IMBH candidates NGC~5286 and NGC~5139 ($\omega$ Centauri) both are more consistent with models that host an IMBH. It is conceivable that clusters with different initial conditions undergo more extreme excursions to low concentrations, and we therefore also considered the clusters with high $r_{10\%}/r_{50\%}$ to the left of the IMBH models as possible candidates. We note that this is almost the entire sample for which surface brightness profiles were available. However, as shown in panel d), some of the clusters are equally well reproduced by a model that contains a high black-hole retention fraction. This degeneracy may be lifted with more detailed $N$-body simulations for both clusters, so that one can distinguish between the two possible explanations. As mentioned before, a black-hole retention fraction of $100\%$, that is, when none of the formed stellar-mass black hole are ejected, does not agree with our knowledge of the violent process of black-hole formation. At least some fraction of the black holes will be ejected because of high kick velocities, and the model with the black-hole retention fraction of $100\%$ should therefore be treated as an extreme case.

To test the results of this study, more kinematic observations have to be performed for the clusters that are considered IMBH candidates. A new sample of inner kinematics for Galactic GCs is desired and planned for the future. A larger sample of GCs with available mass function slopes and surface brightness profiles would also help to better constrain the behavior of GCs in the $\alpha - c$ and $\alpha - r_{10\%}/r_{50\%}$ plane. As we have shown, such observations may provide a powerful diagnostic for identifying GCs that possibly host IMBHs. Furthermore, future $N$-body simulations with an extended parameter space are needed to reproduce the observations that were not covered by the simulations presented in this study.

\begin{acknowledgements}
This research was supported by the DFG cluster of excellence Origin and Structure of the Universe (www.universe-cluster.de). We thank the anonymous referee for constructive comments that helped to improve this manuscript. 
\end{acknowledgements}

\bibliographystyle{aa}
\bibliography{ref}

\end{document}